\documentclass[usenatbib,twocolumn,useAMS]{mn2e}
\usepackage{latexsym}
\usepackage{dcolumn}
\usepackage{bm}
\input{epsf}
\newcommand{\be}{\begin{equation}}
\newcommand{\ee}{\end{equation}}
\newcommand{\ba}{\begin{eqnarray}}
\newcommand{\ea}{\end{eqnarray}}
\newcommand{\bi}{\begin{itemize}}
\newcommand{\ei}{\end{itemize}}
\newcommand{\bfi}{\begin{figure}
\epsfxsize=9cm
\epsffile}
\newcommand{\efi}{\end{figure}}
\newcommand{\mnras}{MNRAS}
\newcommand{\apj}{ApJ}
\newcommand{\apjl}{ApJ}
\newcommand{\apjs}{ApJS}
\newcommand{\aap}{AAP}
\newcommand{\prd}{PRD}

\newcommand{\nat}{Nature}

\title[Photo-z self-calibration]{Self-calibration of photometric redshift
  scatter in weak  lensing surveys}
\author[Zhang, Pen \& Bernstein]{Pengjie Zhang$^1$, Ue-Li Pen$^2$, Gary
  Bernstein$^3$ 
\\$^1$Key Laboratory for Research in Galaxies and Cosmology, Shanghai
  Astronomical Observatory, Nandan Road 80, Shanghai, 200030,
  China;\\ pjzhang@shao.ac.cn\\
$^2$Canadian Institute for Theoretical Astrophysics, University of Toronto, 60 St. George Street, Toronto, ON M5S 3H8, Canada\\$^3$Department of Physics \& Astronomy, University of Pennsylvania, 
Philadelphia, PA 19104, USA}
\begin{document}
\maketitle
\begin{abstract}
Photo-z errors, especially catastrophic errors, are a major uncertainty for
precision weak lensing cosmology. 
We find that the  shear-(galaxy number)
density and 
density-density cross correlation measurements between photo-z bins, available
from the same lensing surveys, contain valuable information for
self-calibration of the scattering probabilities between the true-z and
photo-z bins. The self-calibration
technique  we propose does not rely on cosmological priors nor
parameterization of the photo-z probability distribution function, and preserves
all of the cosmological 
information available from shear-shear
measurement. We
estimate the calibration accuracy through the Fisher matrix formalism. We
find that, for advanced lensing surveys such as the planned stage IV surveys,
the rate of photo-z outliers  can be determined with 
statistical uncertainties of $0.01$-$1\%$ for $z<2$ galaxies.
Among the several sources of
calibration error that we identify and investigate,  the {\it
  galaxy distribution bias} is likely the most dominant systematic error,
whereby photo-z outliers have different redshift distributions and/or bias
than non-outliers from the same 
bin.  This bias affects all photo-z calibration techniques based on correlation
measurements.
Galaxy bias variations of $O(0.1)$ produce biases in photo-z outlier rates
similar to the 
statistical errors of our method, so this galaxy distribution bias may
bias the reconstructed scatters at several-$\sigma$ level, but is unlikely to
completely invalidate the self-calibration technique.  
\end{abstract}

\section{Introduction}

Weak gravitational lensing is emerging as one of the most powerful probes of
dark matter, dark energy \citep{DETF} and the nature of gravity at
cosmological scales \citep{Jain08}.  In less than a decade after first
detections  \citep{Bacon00,Kaiser00,VanWaerbeke00,Wittman00}, the lensing
measurement accuracy and dynamical range have been improved dramatically
(e.g. \citealt{Fu08}).  Future weak lensing surveys have the
potential  to measure the lensing power spectrum with sub-$1\%$ statistical
accuracy for many multipole $\ell$ bins. However, whether we can fully utilize
this  astonishing  capability is 
up to the control over various systematic errors. They could arise from
uncertainties in theoretical modeling, including the non-linear evolution of
the universe \citep{Heitmann05,Coyote1,Coyote2} and the influence of baryons
\citep{White04,Zhan04,Jing06,Rudd08}. They could also arise from uncertainties in the lensing 
measurement. An incomplete list includes
the galaxy intrinsic alignment
\citep{Hirata04,Mandelbaum06,Hirata07,Okumura09a,Okumura09b},  influence of the
telescope PSF \citep{STEP1,STEP2}, photometric redshift (photo-z) calibration
errors \citep{MaHuHuterer, Bernstein09b}, etc. Precision lensing cosmology
puts stringent requirements on calibrating 
these errors \citep{Huterer06}. 

Weak lensing surveys are rich in physics and contain information beyond
the cosmic shear power spectrum \citep{Bernstein09,Zhang08}. This
bonus allows for self-calibration of weak lensing systematic errors,
such as the galaxy intrinsic alignment \citep{Zhang08,Joachimi09}. In the
present paper, 
we utilize the galaxy density-shear cross-correlation and density-density correlations in photometric survey data to self-calibrate the photo-z scatters
between redshift bins. Namely, for a given photo-z bin, we want to figure out
(reconstruct)  the  fraction of galaxies which are actually located in a
distinct redshift bin. These scatters 
quantitatively describe the photo-z outliers or catastrophic errors.  The
effect of photo-z outliers on cosmological inference from the shear-shear
power spectrum is discussed by \citet{Bernstein09b}, which also quantifies the
task of calibrating these outliers by direct spectroscopic sampling of the
galaxy population.  Since spectroscopic sampling of faint galaxies at $>99\%$
completeness is an expensive or infeasible task for current ground-based
capabilities, \citet{Newman08} proposes a technique based on  
cross-correlation between the photo-z sample and an incomplete spectro-z
sample.  

Here we ask a complementary question: how well can the outlier rate be determined using purely photometric data from the original lensing survey?
As pointed out by
\citet{Schneider06}, the spurious cross correlation between the galaxy density in two photo-z
bins can be explored to calibrate
photo-z errors. They found that, for a photo-z bin of the size $\Delta z=0.5$
in a  LSST-like survey, $1\%$ level scatters can be identified and the mean
redshift can be calibrated within the 
accuracy of $\sim 0.01$. This result is impressive. Unfortunately, a factor of
$10$ improvement is still required to meet the statistical accuracy of those
ambitious ``stage IV'' projects 
($\sim 10^{-3}$, \citealt{Huterer06}). In combination with baryon acoustic
oscillation  and weak lensing measurements, the constraints can be
significantly improved \citep{Zhan06b}. However, these procedures adopt a
number of priors/parameterizations, which may bias the calibration.    
 
 A successful self-calibration should not correlate cosmological
uncertainties with astrophysical uncertainties (in our case, photo-z
errors). To meet this requirement, the self-calibration should adopt as few 
cosmological priors as possible, preferably none. On the other hand, it should
not result in loss of cosmological information.  In this paper,
we propose to combine the galaxy-galaxy 
clustering measurement and shear-galaxy cross correlation measurement to
perform the photo-z self-calibration, strictly reserving
the shear-shear measurement for cosmology. Finally, it must be able to reach
sufficiently high 
statistical accuracy and have controllable systematics, if any. As we will
show, this self-calibration 
meets all the three requirements and is able to detect scatters as low as
$0.01\%$. 

There are a number of differences between our self-calibration method and the
method proposed by  \citet{Schneider06}. (1) The inclusion of shear-galaxy cross correlation
measurement breaks a severe degeneracy in the previous method and thus
significantly improves the calibration accuracy.  (2) We do not adopt any
parameterizations on the photo-z probability distribution function (PDF) and
are thus free of possible bias induced by improper parameterizations. (3) This
method is a true self-calibration, in the sense that the photo-z scatters are
reconstructed solely from the given weak lensing surveys, no external
measurements nor priors on cosmology and galaxy bias, are
needed.\footnote{Of course, the calibration accuracy
 depends on the fiducial photo-z PDF or the actual photo-z PDF in the given
 survey.} (4) We argue that the shot noise is the only relevant noise 
term for the likelihood analysis.  Sample variance and non-Gaussianity, do not affect
the reconstruction. This allows us to go deeply into the
nonlinear regime, gain many more independent modes for the
reconstruction, and significantly improve the reconstruction accuracy. 

Like most of these predecessor papers, we analyze an idealized survey:
apparent density fluctuations induced by gravitational lensing are ignored;
galaxy biasing is assumed to be common to all galaxies at a given redshift;
and shear measurement errors are ignored.  Incorporation of these and other
effects can substantially degrade the correlation-based photo-z calibration
methods \citep{Bernstein09b}.  A framework for comprehensive analysis of
photometric$+$spectroscopic lensing survey data is presented in
\citet{Bernstein09}, and is necessary to make a final judgment on the
efficacy of photo-z self-calibration.  However this complicated analysis has
not yet been applied to the problem of photo-z outlier calibration.  The
simpler analysis presented here will demonstrate that the galaxy-shear and
galaxy-galaxy correlations contain sufficient information to measure the
outlier rate to useful precision, and we will then examine the possibility of
degradation by the non-ideal effects. 

This paper is organized in the following way. In \S \ref{sec:calibration}, we
describe our self-calibration technique and target a fiducial ``Stage IV''
lensing survey for the error  
forecast. We discuss in \S\ref{sec:statistical} possible systematic errors
which can be incorporated into our technique and will not bias the photo-z PDF
reconstruction.  
Other systematic errors cannot be self-calibrated without strong priors or external information. For these, we quantify the
induced bias in \S \ref{sec:bias}. We further discuss uncertainties in the
error forecast due to   uncertainties in the fiducial model  and the
robustness of our self-calibration technique (\S \ref{sec:fiducial}). We
discuss  possibilities to improve the
calibration accuracy (\S \ref{sec:discussion}).  We also include two appendices
(\S \ref{sec:appendixA} \& \ref{sec:appendixB})
for technical details of the Fisher matrix analysis and bias estimation.  

\section{Photo-z self-calibration}
\label{sec:calibration}
We first define several key notations used throughout the
paper. 
\bi
\item The superscript ``P'' denotes the property in the photo-z bin.
\item The superscript ``R'' denotes the corresponding property in the true-z
  bin. 
\item The capital ``G'' denotes gravitational lensing, to be more specific,
  the lensing convergence converted from the more direct observable cosmic
  shear.  
\item The little ``g'' denotes galaxy number density (or over-density). 
\ei

We split galaxies into $N_z$ photo-z bins. The $j$-th photo-z bin has the
range $[z_j-\Delta z_j/2,z_j+\Delta z_j/2)$. Our notation is that larger $i$
  means higher photo-z. $N_j$, the total number of
  galaxies in the $j$-th photo-z bin, is an observable. We also have $N_z$
  true-z bins, with the choice of redshift range identical to that of
  the photo-z bins. We denote
  $N_{i\rightarrow j}$ as the 
  total number of galaxies in the $j$-th photo-z bin which also belong to the
  $i$-th true-z bin, namely, whose true redshift fall in $[z_i-\Delta
    z_i/2,z_i+\Delta z_i/2)$. The process
    $i\rightarrow j$ is similar to  scatters (transitions) between  different
    quantum states. So we often call this process  as the photo-z
    scatter. The scatter rate is related to the the  
    photo-z probability distribution function,  $p(z|z^P)$, by the following
    relation 
\be
N_{i\rightarrow j}=\int_{z_j-\frac{\Delta z_j}{2}}^{z_j+\frac{\Delta z_j}{2}}
n(z^P)dz^P\left[\int_{z_i-\frac{\Delta z_i}{2}}^{z_i+\frac{\Delta z_i}{2}} 
p(z|z^P) dz\right]\ .
\ee
Here, $n(z^P)dz^P$ is the number  distribution of galaxies in the photo-z
space.   We define $p_{i\rightarrow j}\equiv N_{i\rightarrow
  j}/N_j$, which represents the 
averaged $p(z|z^P)$ over the relevant redshift ranges, or the
binned photo-z PDF.  In the limit 
that $\Delta z_{i,j}\rightarrow 0$, $p_{i\rightarrow j}\rightarrow
p(z_i|z^P_j)$. Since $\sum_i
p_{i\rightarrow j}=1$,  we have $N_z(N_z-1)$ independent $p_{i\rightarrow
  j}$.\footnote{In general, $p_{i\rightarrow j}$ and $p_{j\rightarrow i}$ are
  independent and $p_{i\rightarrow j}\neq p_{j\rightarrow i}$.}
These $p_{i\rightarrow j}$ together represent a non-parametric
description of the  photo-z PDF $p(z|z^P)$ and completely describe the
scattering probabilities between redshift bins, namely the rate of photo-z
outliers, or leakage rate.   In \citet{Bernstein09b}, they
are called ``contamination'' coefficients.

Scatters in photo-z, especially catastrophic photo-z errors, 
cause a number of spurious correlations between photo-z bins. Our proposal is
to reconstruct all $p_{i\rightarrow j}$ from these cross correlations. For 
convenience, we will work on the corresponding cross power spectra
throughout the paper, instead of the cross correlation functions.

We use fiducial power spectra,
fiducial leakage $p_{i\rightarrow j}$ and the survey specification to perform the
error forecast. To generate the
fiducial power spectra, we adopt a flat
$\Lambda$CDM cosmology with $\Omega_m=0.27$, 
$\Omega_{\Lambda}=1-\Omega_m$, $\Omega_b=0.044$, $\sigma_8=0.84$ and
$h=0.71$. The transfer function is obtained using CMBFAST \citep{CMBFAST}. The nonlinear
matter power spectrum is calculated using the fitting formula of
\citet{Smith03}. Unless specified, we adopt a fiducial galaxy bias $b_g=1$. We
are 
then able to calculate the galaxy power spectra and lensing-galaxy power
spectra. The calculation of these power spectra is by no mean precise, due to
the simplification of $b_g=1$. However, this simple bias model suffices for
the purpose of this paper, namely, to demonstrate the feasibility of our
self-calibration technique. We will further investigate the impact of scale
dependent bias in \S \ref{sec:fiducial}, where we find that our
self-calibration technique is also applicable. 

\bfi{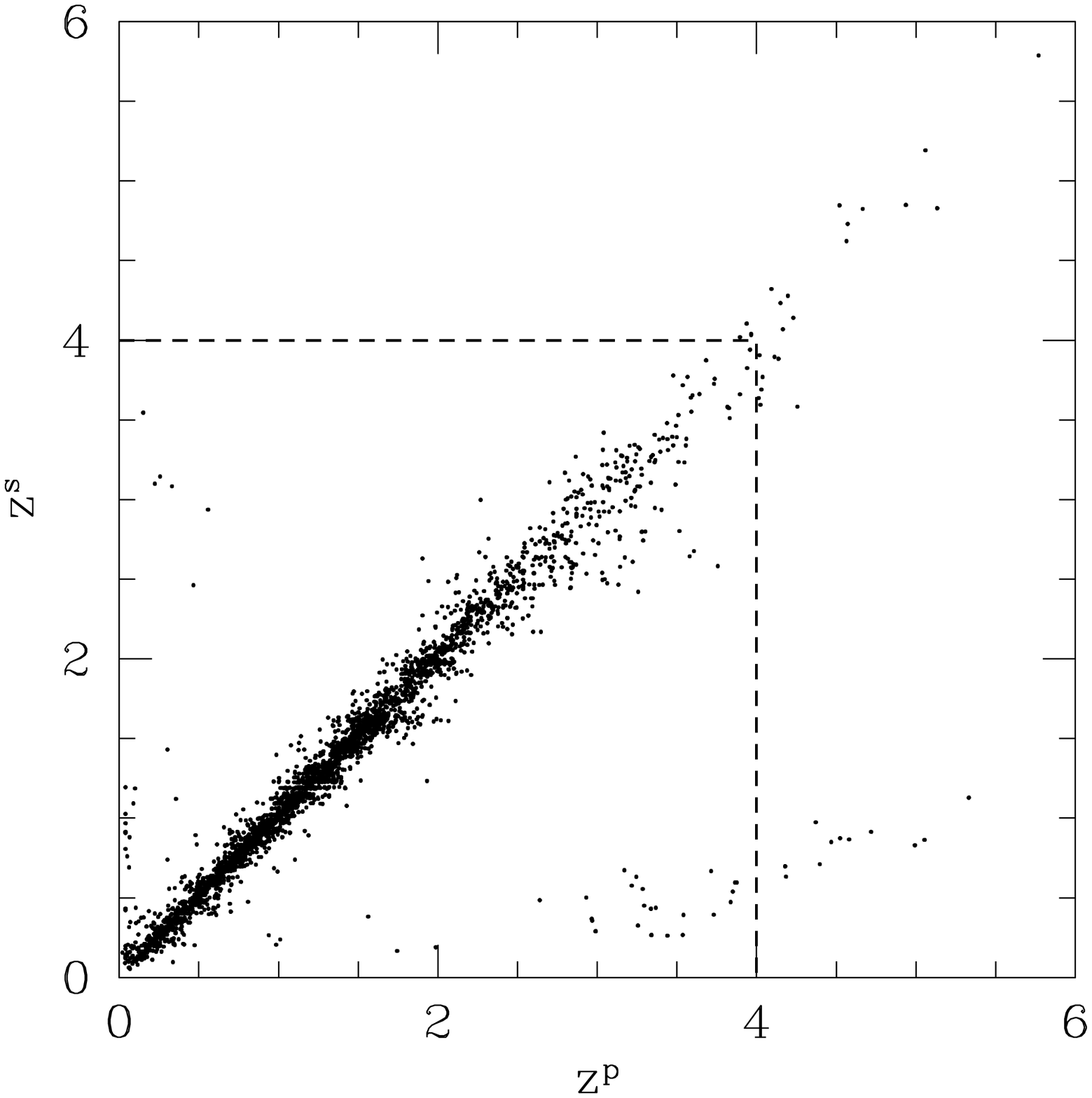}
\caption{The fiducial $z^P$-$z^S$ scatters adopted  for the 
  forecast. The simulated data \citep{Bernstein09b} has $177210$ galaxies. 
To control the size of the figure, we only show $2\%$ of them, randomly
chosen 
Although there are indeed $0.7\%$ galaxies at $z^P>4$, in the
  analysis we have taken the freedom to disregard these galaxies. To make the
  analysis closed,  we shall adopt the approximation that no galaxies with
  $z^P<4.0$ comes from $z^S>4.0$. We check that
  galaxies with $z^S>4.0$ only account 
  for a tiny fraction of the whole $z^P<4$ sample ($0.06\%$) and even smaller
  fraction for low photo-z bins ( and thus not showing up in this figure since
  we only plot $2\%$ of all galaxies). So this approximation 
  is sufficiently accurate for the purpose of this paper.  \label{fig:zszp}} 
\efi

We also need the galaxy distribution $n(z^P)$ and the input of fiducial $p_{i\rightarrow j}$. Due to uncertainties in survey specifications, we will
not focus on any specific survey. Instead, we will target a fiducial
lensing survey with some characteristics  of ``Stage IV'' lensing surveys like
LSST\footnote{Large Synoptic Survey Telescope, http://www.lsst.org/lsst}, Euclid\footnote{http://www.astro.ljmu.ac.uk/~airs2008/docs/euclid\_astronet\%202.pdf} and JDEM. The analysis can be redone straightforwardly to
incorporate changes in the survey specifications.  We follow
\citet{Huterer06,Zhan06} and adopt the photo-z
distribution $n(z^P)dz^P=x^2\exp(-x)dx/2$, where $x\equiv 
z^P/z_0$. For such a distribution, the median redshift is
$z_m=2.675z_0$. We adopt $z_0=0.45$, the mean galaxy surface density 
$\bar{n}_g=40$ per arcmin$^2$, and  the fractional sky coverage $f_{\rm
  sky}=0.5$. The rms dispersion in 
the shear measurement induced by the galaxy intrinsic ellipticities  is
adopted as $\gamma_{\rm rms}=0.2$. 

The fiducial 
$p(z|z^P)$ and $p_{i\rightarrow j}$ are calculated from the simulated data of
\citet{Bernstein09b}, which were produced using the method described in
\citet{Jouvel09}.  The $z^P$-$z^S$ distribution in this data is shown in
Fig. \ref{fig:zszp}, where $z^S$ is the spectroscopic redshift, which can be
approximated as the true redshift. This simulated data is for SNAP-like space
weak lensing  surveys, conducted in 8 broad bands spanning $0.38$-$1.7$$\mu$m wavelength. These surveys have near-IR bands and 
hence likely smaller photo-z 
errors than LSST. Since we find that the statistical accuracy of the
photo-z  reconstruction is not very sensitive to the choice of fiducial
photo-z PDF and since photo-z calibration has room to improve,  the adopted
photo-z PDF should be a good representative case to
illustrate our method.

 Furthermore, the  accuracy of the self-calibration is
sensitive to the choice of redshift bins. Unless otherwise specified, we will
adopt eight redshift bins, $[0.0,0.5)$, $[0.5,1.0)$, $[1.0,1.5)$, $[1.5,2.0)$,
$[2.0,2.5)$, $[2.5,3.0)$, $[3.0,3.5)$ and $[3.5,4.0)$. For comparison, we also
investigate the case of much finer redshift bins, $[0.0,0.1)$, $[0.1,0.2)$,
    $\ldots$, $[1.7,1.9)$, $[1.9,2.2)$, $[2,2,2.6)$, $[2.6,3.0)$, $[3.0,4.0)$. 
We call these coarse bins and fine bins, respectively. We have utilized
the freedom in the data analysis and disregarded
the galaxies with $z^P>4.0$. These galaxies only account for $0.7\%$ of the
total galaxies. Neglecting them does not result in significant loss of
information. On the other hand,  these $z^P>4$ galaxies may have large
photo-z catastrophic errors (Fig. \ref{fig:zszp}). Neglecting them helps to
reduce systematic errors. 

To close the scattering process, we need to assume that no galaxies with
$z^P<4.0$ come from $z^S>4.0$. Under this apssumption, what our
self-calibration technique actually does,   is to assign those $z^S>4$
galaxies into the $[3.5,4.0)$ true-z bin, instead of 
randomly assigning them elsewhere.\footnote{These galaxies do
not lens other galaxies so they have to be put in the highest redshift
bin. More details on how the self-calibration technique rank
  galaxies are given in \S \ref{subsec:fullcalibration}.}  We argue that
  this approximation is 
sufficiently accurate, for two reasons. First, in the simulated $z^S-z^P$
data we used,   the total fraction of galaxies with $z^P<4.0$ and $z^S>4.0$ is
only $0.06\%$.   Second, about $90\%$ of them 
leak into the $[3.5,4.0)$ photo-z bin. Although the  contamination rate in
  this redshift bin is high ($4.8\%$), the induced error in lensing modelling
  is small, since the lensing weighting kernel varies slowly at $z\sim
  4$. Contaminations to other photo-z bins are much smaller. These 
  $z^S>4$ galaxies account for $0.01\%$ of total 
galaxies in the $[0.0,0.5)$ photo-z bin, $0.02\%$  in the $[0.5,1.0)$ and
    $[3.0,3.5)$ photo-z bins.  In the adopted simulated data, we detect no
 such galaxies in photo-z bins $\in [1.0,3.0)$.  Due to these tiny fractions,
mis-assignment of these galaxies is not a limiting factor of our
self-calibration technique.

\subsection{Galaxy-galaxy clustering}
Photo-z errors, especially catastrophic errors, induce non-zero galaxy-galaxy
correlation between different
redshift bins, which should vanish at sufficiently small scale, where the Limber
approximation holds.  This set of correlations  has been explored to perform the
photo-z  self-calibration \citep{Schneider06}.  With the presence of photo-z scatters, the measured
galaxy surface density in a given photo-z bin is the combination of the
corresponding ones in the true redshift bins, 
\be
\label{eqn:g}
\delta_i^{\Sigma,P}=\sum_k p_{k\rightarrow i} \delta^{\Sigma,R}_k\ .
\ee
The galaxy power spectrum between the $i$-th and $j$-th photo-z bins is
\ba
\label{eqn:gg}
C^{gg,P}_{ij}=\sum_{km} p_{k\rightarrow i}p_{m\rightarrow
  j}C_{km}^{gg,R}\simeq \sum_k p_{k\rightarrow i}p_{k\rightarrow j}C_{kk}^{gg,R}\ .
\ea
 The last equation has approximated $C_{k\neq m}^{gg,R}=0$,
which holds under the Limber approximation. At sufficiently large angular
scales, the Limber approximation fails and the intrinsic cross correlation
$C_{k\neq m}^{gg,R}\neq 0$. For this reason, we exclude the modes with 
$\ell<100$. We will further discuss this issue later in the paper (\S \ref{sec:Limber}). 

\bfi{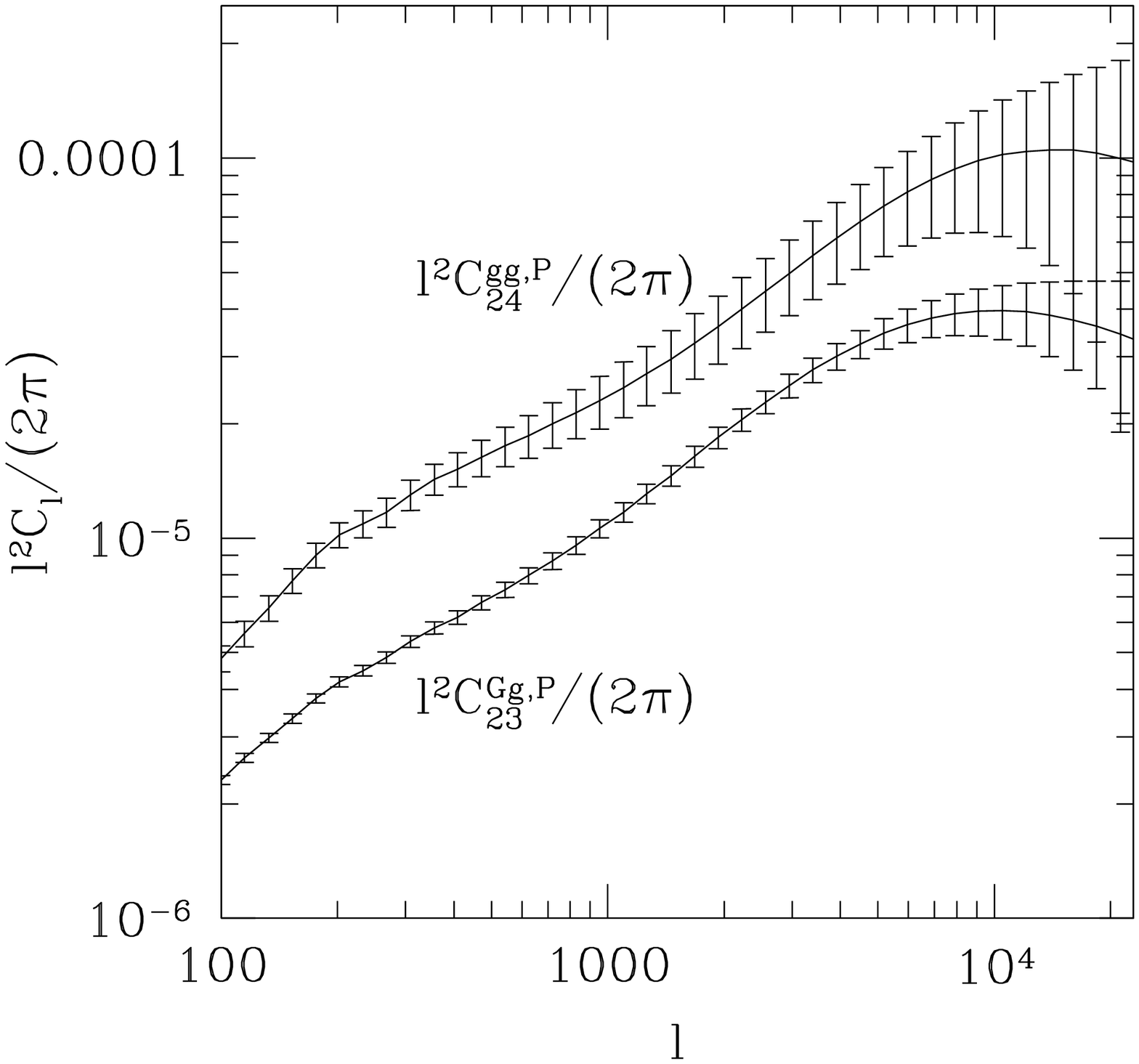}
\caption{Diagnostics of photo-z errors. Scatters between redshift bins caused
  by photo-z errors induce non-zero
  correlations $C^{gg,P}_{i\neq j}$ and  $C^{Gg,P}_{i<j}$, which otherwise
  vanish. Here $i,j=1,2\ldots$ denote different redshift bins, the larger the
  $i,j$, the higher the redshift. These spurious
  cross correlations serve as diagnostics of photo-z 
  scatters. We show the resulting $C^{gg,P}_{24}$ and $C^{Gg,P}_{23}$. The
  errorbars shown are from the shot noise. As explained in the text, this is
  the only relevant source of error.   \label{fig:cl}} 
\efi

If the relevant photo-z scatters are sufficiently large, $C^{gg,P}_{i\neq j}$
will dominate the associated shot noise. In this case, the relevant photo-z
scatters become detectable. For the adjacent bins ($|i-j|=1$), the dominant
contribution to $C^{gg,P}_{ij}$ obviously comes from $p_{i\rightarrow j}$ and
$p_{j\rightarrow i}$, unless the corresponding $p_{i\rightarrow j}$ and
$p_{j\rightarrow i}$ are tiny, which is unlikely. For correlations between
non-adjacent photo-z bins, this may not be true. In
Fig. \ref{fig:cl}, we show the result of $C^{gg,P}_{24}$. In this case, the
dominant contribution comes from $C^{gg,R}_{33}$, through the scatters
$3\rightarrow 2$ and  $3\rightarrow 4$. Due to the huge sky coverage of the
fiducial stage IV lensing survey,
even though the signal $C^{gg,P}_{24}$ is suppressed by a huge factor ($\simeq
p_{3\rightarrow  2}\times p_{3\rightarrow 4}$) relative to $C^{gg,R}_{33}$, it
still overwhelms the shot noise and becomes observable. 

However, an intrinsic degeneracy encoded in the galaxy-galaxy clustering
significantly limits the accuracy of this approach. From
Eq. \ref{eqn:gg}, it is clear that the 
correlation $C^{gg,P}_{i\neq j}$ can be induced by a nonzero
$p_{i\rightarrow j}$ or by a non-zero $p_{j\rightarrow i}$. In other words,
$C^{gg,P}_{ij}(\ell)$ ($i\neq j$) only measures the combination
$p_{j\rightarrow i}p_{j\rightarrow 
  j}C^{gg,R}_{jj}(\ell)+p_{i\rightarrow i}p_{i\rightarrow
  j}C^{gg,R}_{ii}(\ell)$. Thus with 
measurement at a single multipole $\ell$ bin, the galaxy clustering measurement
alone can not break this degeneracy between up and down scatters. Adding more $\ell$ bins can break this
degeneracy---if the ratio $C^{gg,R}_{jj} / C^{gg,R}_{ii}$ varies with
$\ell$. If for example  the 3D galaxy power spectrum is a strict
  power-law over 
the  relevant scales at the relevant redshift range, whose
power index does not vary with redshift,\footnote{This condition is necessary.
  If the power index of the 3D power spectrum varies with redshift, the shape
  of $C^{gg}_{ii}(\ell)$ will vary with  redshift. } then all $C^{gg}_{ii}(\ell)$ are
self-similar and the solution remains
degenerate. Observationally, 
departures from a power law have been  found in the galaxy correlation function
\citep{Zehavi04},  so galaxy clustering
measurements at many $\ell$ bins can break the above degeneracy. Nevertheless,
galaxy correlation functions are observed to be 
close to power laws. This degrades the reconstruction accuracy. Furthermore,
the slope depends weakly on galaxy type.  If one uses
the slight deviations from a power law, small changes in the slope of the
leaking population could lead to large  systematic reconstruction errors. This
is an example of the galaxy distribution bias that we will scrutinize in \S
\ref{sec:galaxybias}.  We will further investigate this issue in \S \ref{sec:fiducial}.

In the next section, we
will show that, due to the unique geometry dependence of gravitational
lensing, the degeneracy between up and down scatters is broken naturally by
adding the lensing-galaxy 
correlation, resulting in significant improvement in the reconstruction.

\subsection{Lensing-galaxy correlations}

Galaxy-galaxy lensing brings $N_z^2$ new observables for each angular scale,
which could be used to break the previous degeneracy. The
scatter can render the otherwise vanishing foreground lensing-background
galaxy cross correlation non-zero ($C^{Gg,P}_{i<j}\neq 0$). More importantly,
lensing, due to its geometry dependence, distinguishes up-scatters from
down-scatters.  This 
new piece of  information is the key to significantly improve the photo-z
self-calibration. 

Without loss of generality, we will work on the lensing convergence $\kappa$,
instead  of the more direct observable cosmic shear $\gamma$, which are
locally equivalent in Fourier space.\footnote{Cosmic shear measurement directly
  measures the reduced shear $\gamma/(1-\kappa)$. So the above statement only
  holds at first order approximation. However, this complexity does not affect
  our self-calibration, in which we do not rely on a theory to predict
  $\gamma$ or $\kappa$.}  With the
presence of photo-z scatters, the measured lensing convergence 
in a given photo-z bin is some linear combination of the ones in true-z bins
weighted by the scatter probability,  
\be
\kappa_i^{P}=\sum_k p_{k\rightarrow i} \kappa^R_k\ . 
\ee
The cross correlation power spectrum between the lensing convergence in the
$i$-th photo-z bin and the galaxy number density in the $j$-th photo-z bin is
given by
\ba
\label{eqn:Gg}
C^{Gg,P}_{ij}&=&\sum_{k\geq m} p_{k\rightarrow i}p_{m\rightarrow
  j}C_{km}^{Gg,R} \ .
\ea 
In the absence of lensing magnification bias, $C_{km}^{Gg,R}\neq 0$ only when the source redshifts are higher than the
galaxy redshifts (namely, $k\geq m$). 
Discussion of magnification bias and other errors will be postponed to
\S\ref{sec:statistical} and \S\ref{sec:bias}.  

We show in Fig. \ref{fig:cl} the case of $C^{Gg,P}_{23}$ as an
example. It is mainly contributed by $C^{Gg,R}_{32}$ through the scatters
$3\rightarrow 2$ and $2\rightarrow 3$, by $C^{Gg,R}_{22}$ from the
scatters $2\rightarrow 3$ and  by  $C^{Gg,R}_{33}$ from the
scattering $3\rightarrow 2$. Given the strength of $p_{3\rightarrow
  2}$ and $p_{2\rightarrow 3}$, the resulting $C^{Gg,P}_{23}$ is sufficiently
large to overwhelm the shot noise at $\ell<10^4$. We notice that, depending on
the values of the relevant $p$ and the size of the redshift bin, the dominant
contribution can come from  the $C^{Gg,R}_{i=j}$ term, despite the heavy
suppression in its amplitude due to the low amplitude of the lensing kernel
over the relevant redshift range.

In most of the cases,  $C^{Gg,P}$ can not be measured with comparable accuracy
to that of  $C^{gg,P}$ (however, refer to Fig. \ref{fig:cl} for one
exception). However,  we point out that it is valuable to include this piece of
information in the photo-z 
self-calibration. It turns out to be the key to breaking the strong degeneracy
between up and down scatters.   Look at the
configuration $i\geq k>j$.  The scatter $i\rightarrow j$ 
contributes to $C^{Gg,P}_{jk}$, while the scatter $j\rightarrow
i$ does  not. For this reason, it can break the degeneracy between
$i\rightarrow j$ and $j\rightarrow i$, encountered in the self-calibration based
on  galaxy clustering alone. The discriminating power relies on the
  intrinsic asymmetry 
  between up and down scatters in generating the lensing effect. So it remains
  efficient, even in the case of self-similar $C^{gg}_{ii}(\ell)$ , where 
  the self-calibration based on  galaxy clustering alone blows up (\S \ref{sec:fiducial}).

\subsection{The photo-z self-calibration combining the galaxy-galaxy and
  lensing-galaxy measurements}
\label{subsec:fullcalibration}
Counting degrees of freedom suggests that 
we should be able to  perform a rather model-independent
self-calibration without any priors. Mathematically, we need to solve Eq. \ref{eqn:gg} \&
\ref{eqn:Gg} for all $p_{j\rightarrow i}$, $C^{gg,R}_{i=j}$ and $C^{Gg,R}_{i\geq j}$
simultaneously. At the beginning of this process,
$C^{gg,P}$ and $C^{Gg,P}$ should be replaced with the corresponding
measurements. The reconstructed power spectra $C^{gg,R}$ and
$C^{Gg,R}$ contain valuable information on cosmology and can be further
explored. The reconstructed
$p_{i\rightarrow j}$ can then be applied to the shear-shear correlation
measurement to correct for bias induced by photo-z
scatters and infer the correct cosmology.  In the present
paper, we will focus on only $p_{i\rightarrow j}$. Thus we treat $C^{gg,P}$
and $C^{Gg,P}$ as nuisance parameters to be  marginalized over.

 The unknown parameters to be determined simultaneously by our
 self-calibration technique are $\lambda=(p_{\mu\rightarrow
  \nu},C^{Gg,R}_{ij}(\ell_1), C^{gg,R}_{kk}(\ell_1)\ldots)$, with $\mu\neq
\nu$, $i\geq j$ and $\mu,\nu,i,j,k=1,\ldots,N_z$.    For $N_{\ell}$ multipole
$\ell$ bins  and $N_z$ 
redshift bins, we have $N_z(N_z+3)N_l/2+N_z(N_z-1)$ quantities to solve ($N_{\ell}N_z$
for $C^{gg,R}_{kk}$, $N_{\ell}N_z(N_z+1)/2$ from $C^{Gg,R}_{km}$ ($k\geq m$) and
$N_z(N_z-1)$ for $p_{k\rightarrow m}$). On the other hand, 
we have $N_{\ell}N_z(3N_z+1)/2$ independent measurements of
correlations. $N_{\ell}N_z^2$ of them come from $C^{Gg,P}_{ij}$ and
$N_{\ell}N_z(N_z+1)/2$ from 
$C^{gg,P}_{i\leq j}$.\footnote{The measurement 
  $C^{gg,P}_{ij}$ is identical to the measurement $C^{gg,P}_{ji}$.}

The equations to solve are quadratic in $p_{i\rightarrow
  j}$ and linear in $C_{ij}$ (Eq. \ref{eqn:gg} 
\& \ref{eqn:Gg}).  For this reason, to guarantee a unique solution for
$p_{i\rightarrow j}$, the number of measurements should be at least
$N_z(N_z-1)$ larger than the number of unknowns. This condition is satisfied
when $N_{\ell}\geq 2$. If all the equations (
  Eq. \ref{eqn:gg} \& \ref{eqn:Gg}) are independent then
  $N_{\ell}=2$ is the minimum requirement. If some of
  the equations are linear combinations of the others, we will need $N_{\ell}>2$. For example, if the galaxy power spectra are strict power laws
  in $\ell$, then even perfect measurements of $C^{gg}(\ell)$ at all $\ell$ and
redshift bins cannot break the degeneracies in $p$ and the self-calibration
fails. In reality, none of the galaxy power spectra and shear-galaxy power
spectra is strictly power law and we expect a valid self-calibration barring another unforeseen degeneracy.  Furthermore, the baryon oscillations leave features in $C^{gg}$
and $C^{Gg}$, which help to improve the reconstruction accuracy
\citep{Zhan06a}.  

To better understand the self-calibration process, we recast
it as a mathematical problem to assign galaxies with photo-z labels into correct
subsets (true-z bins), based on the cross  correlation measurements. The
 condition $C^{gg,R}_{i\neq j}=0$ tells us that these subsets do not overlap with
each other in true redshift. The condition $C^{Gg,R}_{i<j}=0$ sets the
correct order of these true-z bins, ranking from low to high
redshifts, since only galaxies behind a lens can be lensed. Furthermore,
the measured power spectra have different dependences on $C^R$ and
$p_{i\rightarrow j}$ (linearly on $C^{R}$ while 
quadratically on $p_{i\rightarrow j}$). This implies the possibility to
separate $p$ from $C^R$.  

However, careful readers may have already noticed a puzzling behavior. 
Eq. \ref{eqn:gg} \& \ref{eqn:Gg} are invariant under the scaling
\ba
\label{eqn:scaling}
p_{k\rightarrow i}&\rightarrow& f_k^{-1}p_{k\rightarrow i}\nonumber \ ,\\
C^{gg,R}_{kk}    &\rightarrow& f_k^2 C^{gg,R}_{kk}\ , \\
C^{Gg,R}_{km} &\rightarrow& f_kf_m C^{Gg,R}_{km} \ , \nonumber
\ea
where $f_k$ are some arbitrary constants. 
This seems to imply that we are not able to determine $p_{i\rightarrow j}$ without
knowing $C^{gg,R}$ and $C^{Gg,R}$. {\it However, this is not a real
degeneracy}. The reason is that we have the conditions $\sum_{k}
p_{k\rightarrow i}=1$. These $N_z$ constraints uniquely fix the $N_z$ freedom
of $f_k$. In the Fisher matrix analysis carried out in the next section, we
enforce the conditions  $\sum_{k} p_{k\rightarrow i}=1$ by explictly setting
$p_{i\rightarrow i}=1-\sum_{k\neq i}p_{k\rightarrow i}$. 

The above solution to this puzzling problem also leads to the solution of
another question. Can we choose more true-z bins than photo-z bins in the
self-calibration technique? The answer is no. In such
case, the number of unknown constants $f_k$ is larger than the number of
constraints $\sum_{k} p_{k\rightarrow i}=1$. Thus we are not able to uniquely
fix $f_k$ and thus $p_{k\rightarrow i}$. 

The absence of degeneracy in the self-calibration is, in the end, confirmed by
the stability of our Fisher matrix inversion (below) for all the
configurations that 
we have checked, including both the coarse bins and fine bins, various $n(z^P)$ 
and $p(z|z^P)$. \footnote{To be more strict, the
  stability of the Fisher matrix inversion means that the solution is a local
  maximum in the likelihood space, since the Fisher matrix is based on the
  Taylor expansion around the given solution. For the solution to be unique,
  in principle we have to go through the whole likelihood space and prove that
  it is the global maximum. This work is
  beyond the scope of this paper. }

\bfi{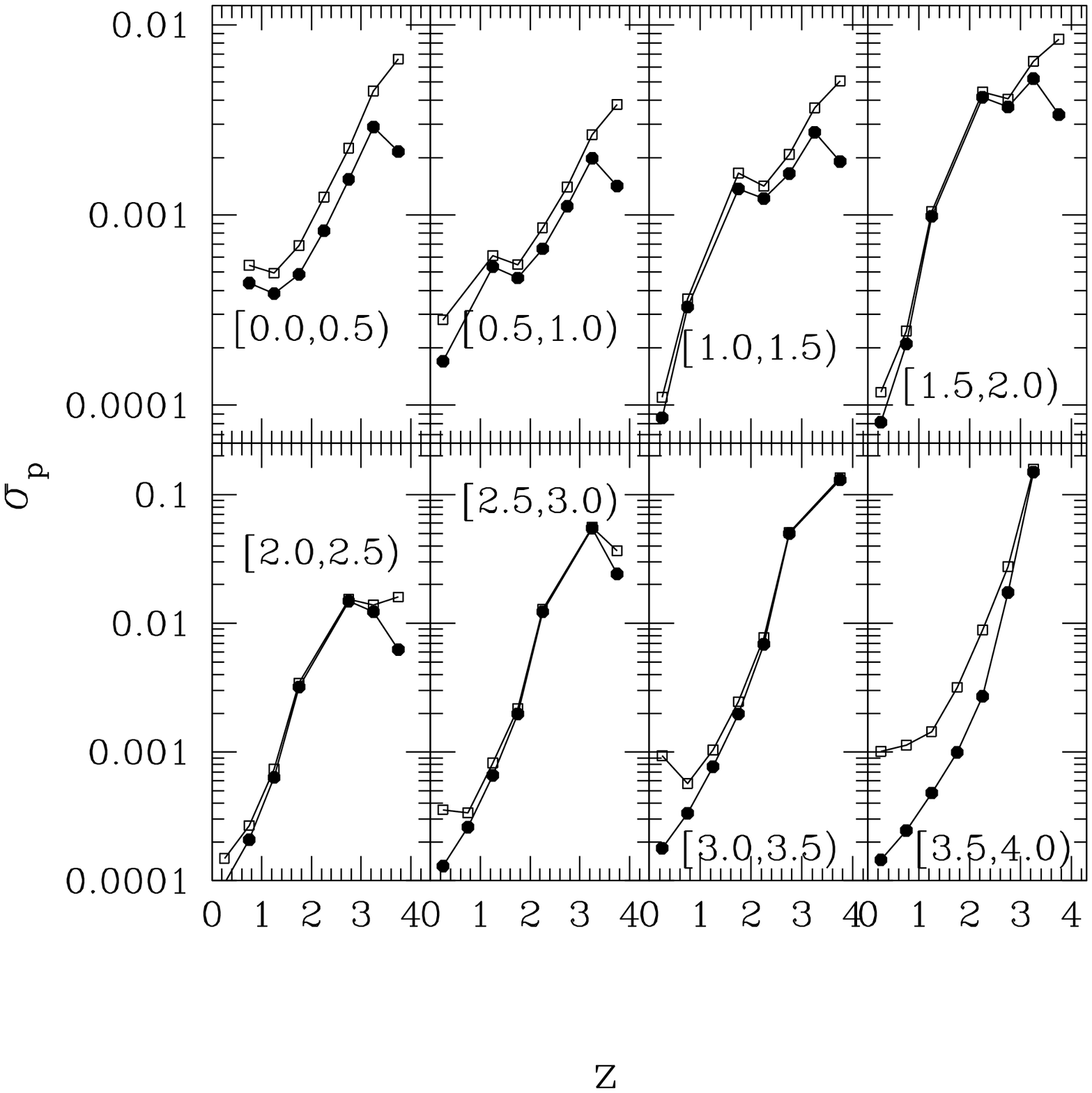}
\caption{The forecasted error in $p_{i\rightarrow j}$ for the fiducial stage
  IV lensing survey. The labels in
  the plots are the photo-z bins. The horizontal axis is the true redshift.
  The filled 
circles are the results including both galaxy-galaxy correlations and
shear-galaxy correlations. The open squares only use information in
galaxy-galaxy correlations. We only show the cases with $i\neq j$, since
$p_{i\rightarrow i}$ is not independent. The horizontal coordinate is $z_i$,
the middle point of the $i$-th photo-z bin, instead of the averaged
true redshift $\langle z_i\rangle$.
\label{fig:p8}} 
\efi

We thus believe that our  photo-z self-calibration does work. It does not rely
on 
any assumption of the underlying photo-z PDF. Furthermore, it does not rely on
cosmological priors, since all cosmology-dependent quantities (e.g. the
galaxy-galaxy and lensing-galaxy power spectra) are self-calibrated
simultaneously. So the reconstructed $p$ is independent of uncertainties
in cosmology.  In the next section, we will quantify the reconstruction error
and show that the proposed self-calibration is indeed powerful. For these
reason, it can be  and should be 
applied to ongoing and proposed weak lensing surveys such as CFHTLS\footnote{http://www.cfht.hawaii.edu/Science/CFHLS/}, 
DES\footnote{https://www.darkenergysurvey.org/}, LSST, JDEM and Euclid.

Before quantifying the reconstruction accuracy, we want to address a
fundamental limitation of this self-calibration technique. It is designed to
diagnose scatters {\em between} redshift bins. It is thus completely blind to
photo-z errors which do not cause such scatter. Any one-to-one mapping
between photo-z and true-z preserves $C^{gg,P}_{i\neq j}=0$ and cannot be discriminated
using galaxy-galaxy correlations.
Some such photo-z errors
can cause $C^{Gg,P}_{i<j}\neq 0$, so we still have some discriminating
power left. If, however, the mapping  between photo-z and true-z is
monotonically increasing, then we have  $C^{gg,P}_{i\neq j}=0$ and
$C^{Gg,P}_{i<j}=0$, and our self-calibration technique completely lacks the
capability to detect such a photo-z error. One simple example is $z^P=(1+\epsilon)z$, where
$\epsilon$ is a constant. We must rely on spectroscopic
redshift measurements to diagnose such errors. In other words, our method can determine only the scattering matrix of photo-z's, and is insensitive to any recalibration of the mean photo-z's.  In this paper, we assume no such mean photo-z
error exists. This is certainly a crucial point of further investigation.

\bfi{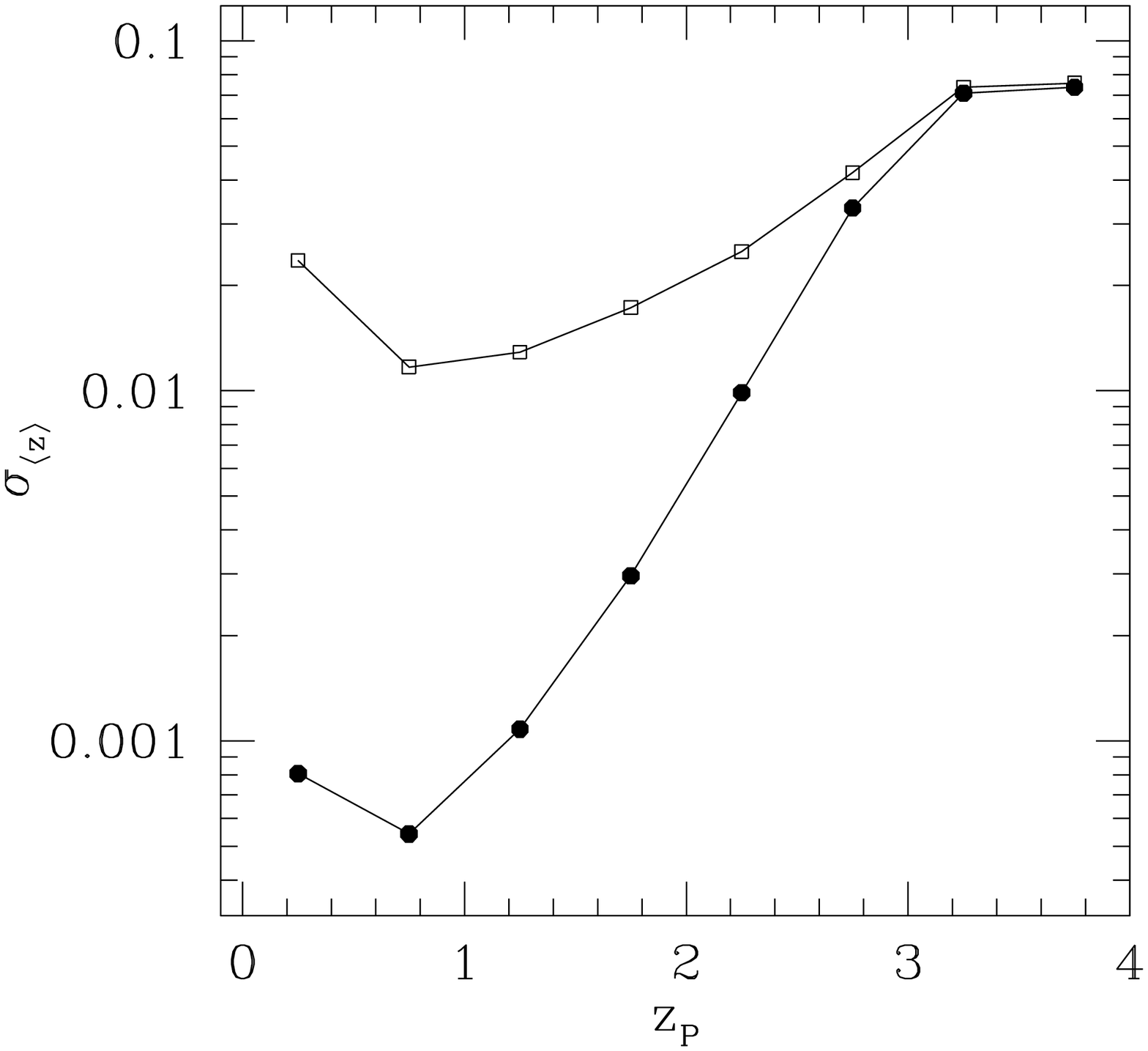}
\caption{The error in the true mean redshift of each photo-z bin for the
  fiducial stage IV lensing survey. The filled 
circles are the results including both galaxy-galaxy correlations and
shear-galaxy correlations. The open squares only use information in
galaxy-galaxy correlations, where we can see the significant improvement by
adding the shear-galaxy measurements. The results are plotted against the
middle-point of the corresponding photo-z bin. We caution that the  galaxy
number density weighted  true-z, $\langle
z_i\rangle$, can differ significantly from the middle-point of the corresponding
photo-z bin, mainly due to significant fractions of scatter into distant
redshift bins.  For example, for the
photo-z bin $[3.5,4.0)$, $\langle z_8\rangle=2.97$, for $[3.0,3.5)$, $\langle
    z_7\rangle=2.7$ and for $[0.0,0.5)$, $\langle z_1\rangle=0.426$.  \label{fig:zmean}}
\efi

\subsection{Error estimation}
\label{subsec:error}
We derive the likelihood function and adopt the Fisher matrix formalism to
estimate the capability of our self-calibration technique. The details are presented in the appendices
\ref{sec:appendixA} \& \ref{sec:appendixB}. We want to highlight that the
error estimation here is 
distinctly  different  from that in routine  
exercises of cosmological parameter constraints and that in
\citet{Schneider06}. In these cases, the theory 
predicts the ensemble average power spectra, which are then compared to the
data. An 
inevitable consequence of any power-spectrum determination is uncertainty due
to cosmic (sample) variance.  But in our self-calibration, the cosmic variance
does not work in 
this way, because the $C^R$ are fitted parameters, not theoretical
predictions. What enters into the key equations \ref{eqn:gg} and 
\ref{eqn:Gg} is not the ensemble average of the power spectra, but the actual
values in the observed cosmic volume, {\it i.e.} they are the sums of their
ensemble averages and cosmic variances within the observed volume. Galaxies in
the same true-z bin (but different photo-z bins) share the same cosmic volume,
thus their power spectra share the same sample variance (however, see \S
\ref{sec:galaxybias}  for complexities), as do their cross
power spectra with the matter. Such coherence has been pointed out by
\citet{Pen04} and has been applied to improve the weak lensing measurement
\citep{Pen04} and primordial non-Gaussianity measurement through two-point
galaxy clustering \citep{Seljak09}. Furthermore, 
we do not rely on a  
cosmological theory to predict these power spectra.   Instead, our self-calibration  reconstructs the actual power spectra  in the
observed survey volume. For this reason,  the only relevant source of
noise in writing down the  likelihood function  is the shot
noise.\footnote{We also want to address that 
  this does not mean that the influence of cosmic variance 
vanishes magically in the Fisher matrix error forecast. In fact, {\it it
  enters the fiducial power spectra}, since 
the fiducial ones should be those measured in a given cosmic volume instead
of the ensemble average. To carry out 
the Fisher matrix analysis more robustly, we need to generate many
realizations of the fiducial power spectra, do the Fisher matrix analysis, and
weigh the error forecast according to the probability of each realization of
the fiducial power spectra to find out the final answer. The good thing is
that the cosmic variance of each power spectrum is usually much smaller than
the ensemble average ($\ell \Delta lf_{\rm sky}\gg 1$), given the
large sky  coverage of the fiducial stage IV lensing survey. The
reconstruction error in our self-calibration is not 
sensitive to such small fluctuations in the fiducial power spectra. Thus, we
are safe to skip the full process and just use the ensemble average as the
fiducial  power spectra. For forecasts of surveys with  much smaller sky
coverage, such as  CFHTLS, we may need to go through the full
process. Furthermore, if we want to infer cosmology from  the reconstructed
power spectra, cosmic variance definitely enters.  }

\bfi{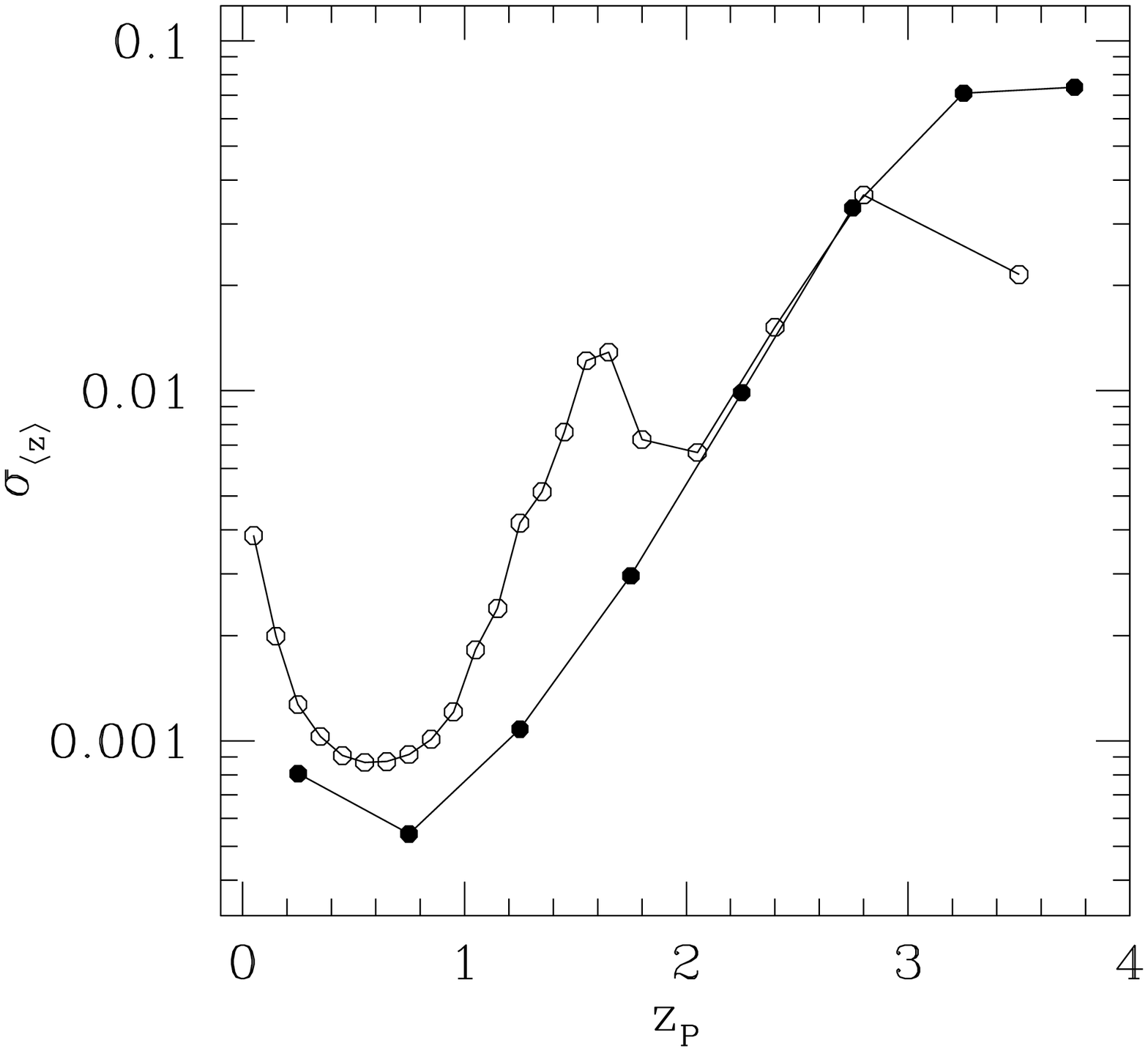}
\caption{The dependence of reconstruction error on the size of redshift
  bins. The  filled 
circles are the results showing in the previous figure adopting coarse
redshift bins and the open circles adopt finer redshift bins. Changes in
redshift bin size not only change the level of noise and the strength of
signal, but also the number of parameters to be fitted. Along with the
complicated error propagation, we see a quite irregular dependence on the bin
size. 
 \label{fig:zmean22}}
\efi

This point is of crucial importance for our error analysis. (1) It allows us
to derive the likelihood function robustly in essentially all $\ell$ range. It
is simply Gaussian, thanks to the central limit theorem and the stochasticity
of shot noise. This is even true for the high-$\ell$ regime, where the
underlying 
density fields are highly nonlinear and non-Gaussian, but the shot noise remains
Gaussian over many independent $\ell$ modes.  (2) Since we do not rely upon any theoretical model for the power spectra, we do not need a theory capable of predictions at small scales where non-linear and baryonic physics are important.
(3) For these reasons, we do not need to disregard those $\ell$ measurements in
highly nonlinear regime, as \citet{Schneider06} did. The inclusion of these
measurements significantly improves the $p$ reconstruction, especially for low
redshift bins. This explains much of the difference between the reconstruction
errors of this paper from that of \citet{Schneider06}, using the galaxy
clustering measurement alone. The high-$\ell$ limit of this analysis will
ultimately depend upon other limitations such as the applicability of the
weak-lensing approximation at small scales, which can in principle render
  the adopted fiducial power spectra unrealistic and thus the error forecast
  unrealistic. However, in our exercise, we expect and have numerically
  confirmed that 
the
contribution of high-$\ell$ is highly suppressed by the shot noise (as can be
seen from Fig. \ref{fig:cl}), so the reconstruction accuracy is not sensitive to
the high-$\ell$ limit. Throughout this
paper, the results shown are based on the choice of $100<\ell<10^{5}$. 
(4) It significantly simplifies the matrix
inversion and improves the numerical accuracy. For $N_z\sim 10$ and $N_l\sim
100$, the Fisher matrix to invert is several   thousand by several
thousand. However, a dominant portion of this Fisher matrix is block 
diagonal, due to the shot noise feature of the error sources. This allows us
to significantly reduce the work on matrix 
inversion. The detail is explained in the appendix.

\bfi{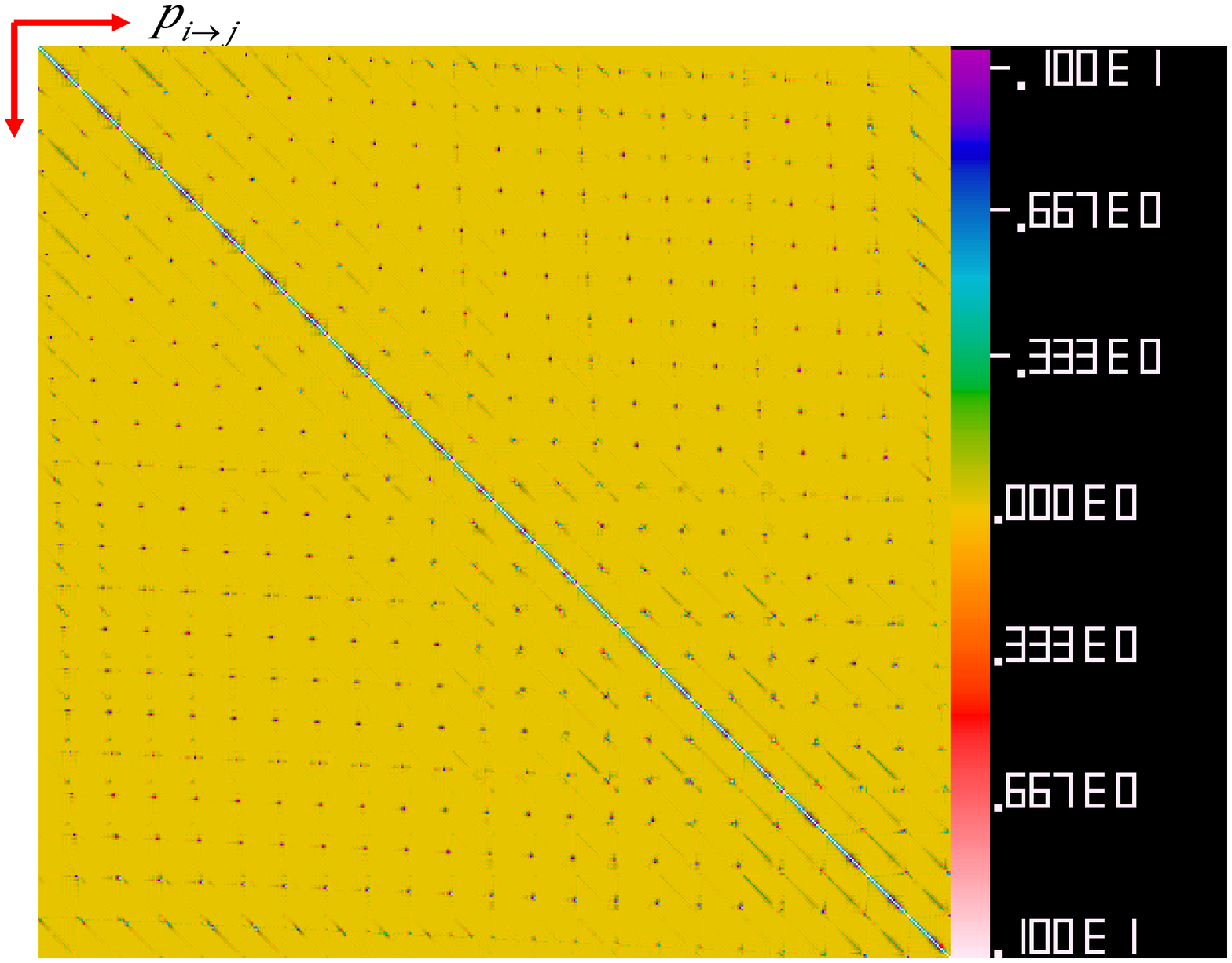}
\caption{The cross correlation coefficient between errors in $p$. To
  demonstrate the fine structure, we choose the fine redshift  bins ($N_z=22$
  redshift bins  in total). Both the horizontal and vertical axes are  $p_{i\rightarrow j}$
  ($i\neq j$), which runs with the order
  $(i,j)=(2,1),(3,1),\ldots,(N_z,1),(1,2),(3,2)\ldots, (N_z,2),\ldots$. There
  are 462 $p$s in total. Errors
  in $p_{i\rightarrow j}$ and $p_{k\rightarrow j}$ are correlated, since they
  could cause some $C^{Gg,P}\neq 0$. Errors in $p_{i\rightarrow j}$ and
  $p_{i\rightarrow k}$ are correlated too, since they cause $C^{gg,P}_{jk}\neq
  0$. More fine structure can be found by zooming in this
    figure. \label{fig:r}} 
\efi

We show the error forecast in Figs. \ref{fig:p8}, \ref{fig:zmean},
\ref{fig:zmean22} \& \ref{fig:r}. For most
bins at $z<2$, $p$ can be
reconstructed to accuracy $<0.01$, for either coarse or fine bins. 
The improvement by adding the
shear-galaxy measurement is often better than $30\%$, up to a factor
of a few. We can also compress
errors in  $p_{i\rightarrow j}$ into a single number $\sigma_{\langle
  z\rangle}$, the  statistical error in the mean true redshift of each photo-z
bin. $\sigma_{\langle z\rangle}$ by no means captures all information of the
reconstruction error, but it is a convenient reference. The result of
$\sigma_{\langle z\rangle}$ for coarse bins is shown in Fig. \ref{fig:zmean}
and that for fine bins is shown in Fig. \ref{fig:zmean22}. For the coarse bins,
it can reach $O(10^{-3})$ for $z<2$. The improvement by adding the 
lensing-galaxy measurement is a factor of $\sim 10$ at $z<2$, mainly through
the improvement in constraining scatters from high redshift
bins. $\sigma_{\langle  
  z\rangle}$ for fine bins is larger (Fig. \ref{fig:zmean22}). However, these
errors are tightly  anti-correlated, as can be inferred from
Fig. \ref{fig:zmean22} and \ref{fig:r}. This is the reason we see big
improvement when choosing bigger bin size.

We present an order of magnitude estimation to understand these numbers of the
reconstruction error. For example, many $p_{i\rightarrow j}$ can be determined
within the accuracy of 
$O(10^{-4})$. We take ${p_{2\rightarrow 1}}$ as an example. The scatter
$2\rightarrow   1$ causes $C^{gg,P}_{12}\neq 0$. Ignoring other scatters, the
threshold of  $p_{2\rightarrow 1}$ is  roughly set when  the accumulated
signal-to-noise of the $C^{gg,P}_{12}$ measurement is 1,
\be
\left(\frac{S}{N}\right)_{12}^2\sim \sum_{\ell}
\left(\frac{C^{gg,P}_{12}}{\sigma^{gg}_{12}}\right)^2=p_{2\rightarrow 1}^2\sum_{\ell}
\left(\frac{C^{gg,R}_{22}}{\sigma^{gg}_{12}}\right)^2=1\ .
\ee
Here, $\sigma$ is the associated shot noise power spectrum. We find that the
threshold $p_{2\rightarrow 1}$ inferred from the above equation is even smaller
than  $\sigma_{p_{2\rightarrow 1}}$, the  statistical error in 
  $p_{2\rightarrow 1}$ (Fig. \ref{fig:p8}).  This is indeed expected. Due to
simplifications made in the derivation, mainly the neglect of error
propagation from  other $p_{i\rightarrow j}$ and $C_{ij}$, the threshold of
$p_{2\rightarrow 1}$  obtained from the above approximation is certainly a
lower limit of $\sigma_{p_{2\rightarrow 1}}$.  

From similar arguments, we also
expect that the statistical errors for $p_{i\rightarrow j}$ are large for
those high redshift bins (e.g. $i=7,8$), because the number of galaxies in
these high redshift bins is small and thus the shot noise is large.
However, as 
explained early, $C^{gg,P}$ is not the only source of information for the $p$
reconstruction. $C^{Gg,P}$ can also play important role. For example, the
reconstruction of $p_{1\rightarrow 8}$ can reach an accuracy of
$10^{-4}$. The reason is that the scatter $1\rightarrow 8$ causes
$C^{Gg,P}_{1j}\neq 0$. The combined  G-g measurements have $S/N>1$ even for
$p_{1\rightarrow 8}=10^{-4}$. The $C^{gg,P}_{18}$
measurement also contributes, but since the number of galaxies in the
$8$-th bin is only 1\% of total galaxies, the associated shot noise is
large. So its contribution is overwhelmed by that from $C^{Gg,P}_{1j}$
($j<8$). This explains the factor of 10 improvement in the $p_{1\rightarrow 8}$
reconstruction when adding the G-g measurements. Finally, we caution that,
although we are able to  qualitatively explain some results of
Fig. \ref{fig:p8}, the error prorogation is complicated and the above
estimation only serves as a convenient tool to understand Fig. \ref{fig:p8}. 

$\delta p_{i\rightarrow j}$, the errors in $p_{i\rightarrow j}$,  are
correlated, and the correlations show rich 
structures. To better demonstrate these features, we adopt the fine
redshift bins, $N_z=22$ in total. We define the cross correlation coefficient
between $\delta p$, as $r_{i_1j_1;i_2j_2}\equiv \langle \delta p_{i_1\rightarrow j_1}\delta
p_{i_2\rightarrow j_2}\rangle/\sqrt{\langle \delta p_{i_1\rightarrow
  j_1}^2\rangle\langle \delta 
p_{i_2\rightarrow j_2}^2\rangle}$. The resulting $r$ is shown in Fig. \ref{fig:r}. Strong positive and negative correlations
exist for errors between many scatters into the same photo-z bins
($p_{i\rightarrow j}$ and $p_{k\rightarrow j}$, regions around the diagonal of
Fig. \ref{fig:r}). These scatters are coupled since both reduce
$C^{Gg}_{j<m}\neq 0$ ($i>m$ and $k>m$) or $C^{Gg}_{m<j}$ ($m>i$ and
$m>k$). Scatters $i\rightarrow j$ and $i\rightarrow k$ are coupled, too, since
they contribute to $C^{gg,P}_{jk}\neq 0$. This explains some strong (both
positive and negative)  correlations of the off-diagonal elements.

\section{Extra sources of statistical errors}
\label{sec:statistical}
Our self-calibration technique does not rely on priors on cosmology or photo-z
distribution. In this sense, it is robust. However, there are still several
sources of systematic error. Some of them, if handled properly, can be rendered
into statistical errors, without resorting to external information,
and will not bias our reconstruction of
$p_{i\rightarrow j}$. We will discuss them in this section. The
remaining of them can not be incorporated into the self-calibration without
strong  priors and will be discussed in \S \ref{sec:bias}.

We find that galaxy intrinsic alignment (\S \ref{sec:IA}), the magnification
and  size bias (\S \ref{sec:magnification}) and the intrinsic cross correlation
between different galaxy bins (\S \ref{sec:Limber}), can in principle be
incorporated into 
our  self-calibration technique and thus do not bias the $p$ reconstruction.
Furthermore, we argue that 
the  inclusion of these complexities is unlikely to significantly
degrade the accuracy of our self-calibration technique.

\subsection{The intrinsic alignment}
\label{sec:IA}
Surprisingly, galaxy intrinsic alignments do not bias the 
$p_{i\rightarrow j}$ reconstructed through our self-calibration technique for
$p_{i\rightarrow j}$, although they definitely bias the inferred $C^{Gg}$ values. 
With the presence of the intrinsic alignment $I$, Eq. \ref{eqn:Gg} becomes
\ba
C^{Gg,P}_{ij}&=&\sum_{k>m}p_{k\rightarrow i}p_{m\rightarrow
  j}C^{Gg,R}_{km}\nonumber \\
   &+&\sum_k p_{k\rightarrow i}p_{k\rightarrow
  j}\left[C^{Ig,R}_{kk}+C^{Gg,R}_{kk}\right]\ .
\ea
Since we do not make any assumption on $C^{Gg,R}_{kk}$, our self-calibration
technique automatically takes 
the intrinsic alignment into account and measures the sum of $C^{Gg,R}_{kk}$ and
$C^{Ig,R}_{kk}$. Clearly, it does not bias the
reconstruction of $p$. On the other hand, it certainly affects the
statistical accuracy of the reconstruction of $p$. Unless
$|C^{Ig,R}_{ii}|\gg C^{Gg,R}_{ii}$,  its existence does not affect the error
forecast significantly. 

If the intrinsic alignment $C^{Ig,R}_{kk}$ depends upon galaxy properties, it is possible
that this term will differ for outlier galaxies than for those correctly assigned to photo-z
bin $k$.  In this case a bias in $p_{k\rightarrow i}$ may result.  This behavior is similar to
the systematics from variation of $b_g$ that are discussed in more detail in 
\S\ref{sec:galaxybias}.

\subsection{Magnification and size bias}
\label{sec:magnification}
In reality, the measured galaxy distribution is the one lensed by foreground
matter distribution. The measured galaxy over-density then has extra
contribution from the lensing. Besides the well known magnification bias due
to the lensing magnification on galaxy flux, there is also a size bias due to
the lensing magnification on galaxy size \citep{Jain02,Schmidt09}. Both can be
incorporated into a function 
$g(F,A)$, determined by the flux $F$ and size $A$ distribution of galaxies in
the  
given redshift bin. The lensed galaxy over-density then takes the form
$\delta_g\rightarrow \delta_g+g(F,A)\kappa$.  

The existence of this extra term induces non-vanishing $C^{gg,R}_{i\neq j}$
and $C^{Gg,R}_{i<j}$. If not taken into account, it will certainly bias the
$p_{i\rightarrow j}$ reconstruction. The good thing is, at least in principle,
the same weak lensing surveys contain the right information to correct for
this effect. Given a lensing survey, we are able to split galaxies into bins
of flux and size. Since the prefactor $g$ is determined by the flux and size
distribution and is a measurable quantity, we are able to separate its effect
from others. Or, alternatively, we can design an estimator
$W(F,A)$ such that $\langle gW\rangle=0$, averaged over all flux and size
bins. The price to pay is the statistical accuracy of the correlation
measurement. For galaxy clustering, the shot noise increases by a factor
$\langle W^2\rangle\langle b_g\rangle^2/\langle Wb_g\rangle^2$, with respect
to the clustering signal. Here,  $b_g(F,A)$ is the galaxy bias.  Robust
modeling of this factor requires information on 
galaxies to high redshifts and faint luminosities.
Furthermore, we need to evaluate the effect of measurement error on
galaxy flux and size. None of these exercises are trivial, so we postpone
such studies elsewhere.  However, we argue qualitatively that the
degradation  in the statistical accuracy is not likely dramatic. Since $b_g$
is always positive and $g$ changes sign from 
the bright end down to the faint end of the galaxy luminosity function, we do not
expect a large loss of statistical accuracy by such weighting.  However the
success of the self-calibration will depend upon the accuracy of methods to
estimate $g(F,A)$ (see also \citealt{Bernstein09b}).

\subsection{The intrinsic galaxy cross correlation between non-overlapping
  redshift bins}
\label{sec:Limber}
 Under the Limber approximation, the galaxy cross correlation between
non-overlapping redshift bins vanishes. However, the Limber approximation is
not $100\%$ accurate. In reality, there is indeed a non-vanishing intrinsic
galaxy cross correlation  $C^{gg,R}_{i\neq j}\neq 0$, especially at large
scales. 
As correctly pointed out by \citet{Schneider06},  this intrinsic cross
correlation  biases the reconstruction of $p$. Eq. \ref{eqn:gg} should  now be
replaced by  
\ba 
\label{eqn:Limber}
C^{gg,P}_{ij}&=&\sum_{k}p_{k\rightarrow i}p_{k\rightarrow j}
C^{gg,R}_{kk}+\sum_{k\neq m}p_{k\rightarrow i}p_{m\rightarrow j}
C^{gg,R}_{km}\\ 
&\simeq & \sum_{k}p_{k\rightarrow i}p_{k\rightarrow j}
C^{gg,R}_{kk}+\sum_{k-m=\pm 1}p_{k\rightarrow i}p_{m\rightarrow j}
C^{gg,R}_{km} \ . \nonumber
\ea 
In the last expression, we only keep the correlation between two adjacent
redshift bins and neglect the correlations between non-adjacent redshift bins
($C^{gg,R}_{|k-m|>1}$). This approximation should be 
sufficiently accurate in practical applications.  Thus even if no photo-z
error is presented, there is still an intrinsic (real) 
correlation between two different (especially adjacent) redshift bins. If not
accounted for, these non-zero $C^{gg,R}_{k\neq m}$ will be mis-interpreted as a
photo-z error and thus bias the reconstruction of $p$.

We first attempt to quantify $C^{gg,R}_{k\neq m}$. Since we only need to
evaluate $C^{gg}_{k\neq m}$ at large scales where it is relevant, we can adopt
the   linear theory to calculate it through the following well known formula 
\be
C^{gg,R}_{ij}=\int_0^{\infty} \Delta^2_m(k,z=0)\frac{dk}{k}Q_i(k,l)Q_j(k,l)\ ,
\ee
where
\ba
Q_i(k,l)=\frac{\int_{z_i-\Delta z_i/2}^{z_i+\Delta z_i/2}
D(z)b_g(z)j_l(k\chi)n(z)dz}{\int_{z_i-\Delta z_i/2}^{z_i+\Delta z_i/2}
  n(z)dz}\ .
\ea
Unfortunately, since $Q_iQ_j$ ($i\neq j$) oscillates around zero and positive
and negative contributions to the integral largely cancel, the
numerical integration is very sensitive to numerical errors and is thus highly
unstable. Nevertheless, through the Monte Carlo numerical integral, we believe
that, at $\ell\geq 100$, the cross power spectrum  between redshift bins
$[0.0,0.5)$ and $[0.5,1.0)$ falls below $\sim 1\%$ of the 
geometrical mean of the corresponding two auto correlation power spectra,
confirming the findings of  \citet{Schneider06}.
More accurate evaluation of the intrinsic cross
correlation may be performed in real space, where we can avoid the highly
oscillating integrand encountered in the multipole space. This issue will be
further investigated. 

 The bias induced is roughly $\delta p_{k\rightarrow m}\sim
 C^{gg,R}_{km}/C^{gg,R}_{kk}$ 
  or $\delta p_{m\rightarrow k}\sim C^{gg,R}_{km}/C^{gg,R}_{mm}$. Depending on the
  low $\ell$ cut, this bias may become comparable to the statistical accuracy
  of self-calibration method. However, there are several possible ways to
  eliminate or reduce this bias. 

One way is to start with the last approximation of Eq. \ref{eqn:Limber}, treat
$C^{gg,R}_{k-m=\pm 1}$ as free parameters and fit them simultaneously with
other parameters. This can eliminate virtually all the associated bias, with
the expense of larger statistical errors. We can easily figure out that there
are still many more  measurements than unknowns, so this remedy is
doable. Furthermore, the degeneracy between  $C^{gg,R}_{k-m=\pm 1}$ and $p$ is
weak. For example, $C^{gg,P}_{i\neq j}$ induced by the scattering
$p_{i\rightarrow j}$ has the property $ C^{gg,P}_{i\neq
  j}/C^{gg,R}_{ii}\propto \ell^0$. On the other hand, the intrinsic cross
correlation induced by the deviation from the Limber approximation decreases
quickly with $\ell$ and thus $C^{gg,R}_{i\neq j}/C^{gg,R}_{ii}$ decreases
quickly with $\ell$. These distinctive behaviors help to distinguish  the
intrinsic cross correlation from the one induced by photo-z errors.  The
characteristic behavior of $C^{gg,R}_{i\neq j}/C^{gg,R}_{ii}$ allows us to
take priors which  are weak, while still helpful to discriminate between
$C^{gg,R}_{i\neq j}/C^{gg,R}_{ii}$ and $p$. For example, we can set
$C^{gg,R}_{i\neq j}=0$ when 
$\ell>\ell_{\rm Limber}$ and thus reduces the number of extra
unknowns. Alternatively, we can model $C^{gg,R}_{i\neq j}/C^{gg,R}_{ii}$ as a
power law of decreasing power with respect to $\ell$. 

The inclusion of the lensing-galaxy cross correlation measurement also helps
to break the degeneracy between $C^{gg,R}_{i\neq j}\neq 0$ and $p$. The
photo-z scatters induce both non-zero $C^{gg,P}_{k\neq m}$ and
$C^{Gg}_{i<j}$. On the other hand, the failure of the Limber approximation
does not cause   $C^{Gg,R}_{i<j}\neq 0$, since the lensing kernel vanishes. 

We then conclude that the intrinsic galaxy cross correlation between
non-overlapping redshift bins may
be non-negligible for stage IV lensing surveys. However,
our self-calibration technique has the capability to take this complexity into
account. Further investigation is required to quantify its influence on the
self-calibration.

\section{Possible systematics}
\label{sec:bias}
There are some error sources which cannot be incorporated into our
self-calibration technique without strong priors or without external
information. Thus they will bias the reconstructed photo-z scatters.  We
discuss the influence of the {\it galaxy distribution bias} 
in \S \ref{sec:galaxybias} and the {\it multiplicative error bias} in cosmic
shear  measurement \S \ref{sec:f}.

\bfi{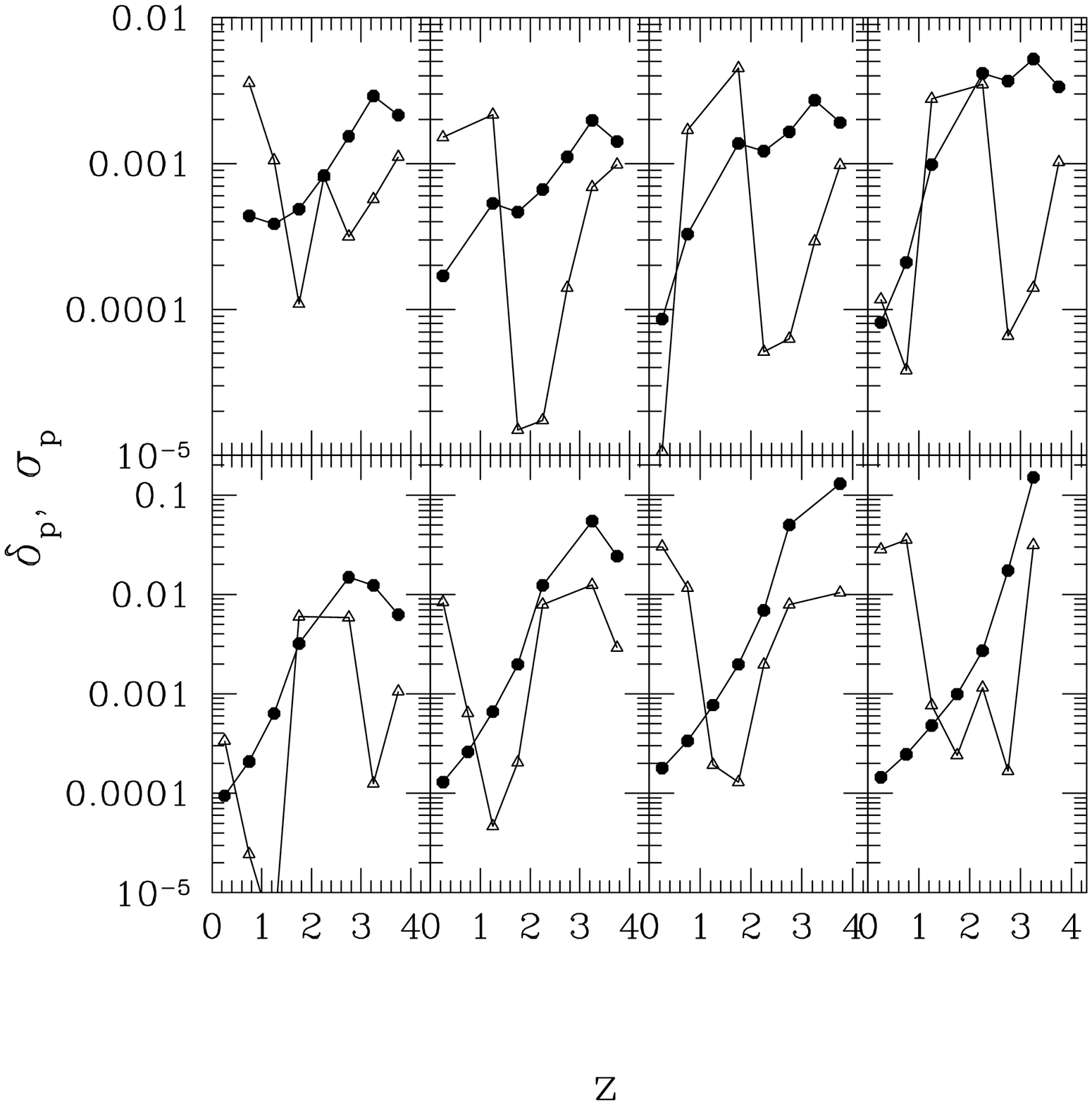}
\caption{The systematic error induced in $p$ by the galaxy distribution bias
  $\tilde{b}$. Circles: statistical errors. Triangles: systematic
  errors (the absolute values). These are for LSST. For DES and CFHTLS, the systematic errors
  induced by the relative galaxy bias is sub-dominant and negligible. The toy
  model adopt for $\tilde{b}$ takes the form $\tilde{b}_{i\rightarrow
    j}=1+s(z_i-z_j)$ and we adopt $|s|=0.1$ for the result shown here. The
  induced bias scales as $s$.   \label{fig:bias}}
\efi

\subsection{The galaxy distribution bias}
\label{sec:galaxybias}
A crucial assumption in the existing self-calibration technique is that those
galaxies scattering
out of the true redshift bin have the same spatial distribution as those
remain in the true redshift bin.  This implicit assumption can be inferred
from Eq. \ref{eqn:g}. By straightforward math, we can find that
$\delta_{k}^{\Sigma, R}$ in this equation is actually
\be
\delta_{k,i}^{\Sigma, R}=\frac{\int_k n_i(z) \delta_{g,i}(z)dz}{\int_k n_i(z)dz}\ ,
\ee
where the integral is over the $k$-th redshift bin, $n_i(z)=\int_i
p(z|z^P)n(z^P)dz^P$ is the true 
redshift distribution of the $i$-th photo z bin,  and $\delta_{g,i}$ is the overdensity
for galaxies in this photo-z bin. In Eq. \ref{eqn:g}, there is
an implicit approximation $\delta_{k,i}^{\Sigma, R}=\delta_{k,k}^{\Sigma,
  R}=\delta_{k}^{\Sigma, R}$.  

This assumption is likely problematic. Those galaxies scattering from true
redshift bin $k$ to photo-z bin $i$ could have either different 
redshift distribution ($n(z)$) or different clustering ($b_g$) or both
compared to galaxies correctly identified in photo-z bin $k$. 
Furthermore, the difference in $n(z)$ means that
these subcategories of galaxies do not sample the cosmic volume with identical
weighting (despite sharing the same true-z bin), so they do not share exactly
the same cosmic variance and thus do not have the identical clustering
pattern. We call all these complexities as the {\it galaxy distribution
  bias}. 

\bfi{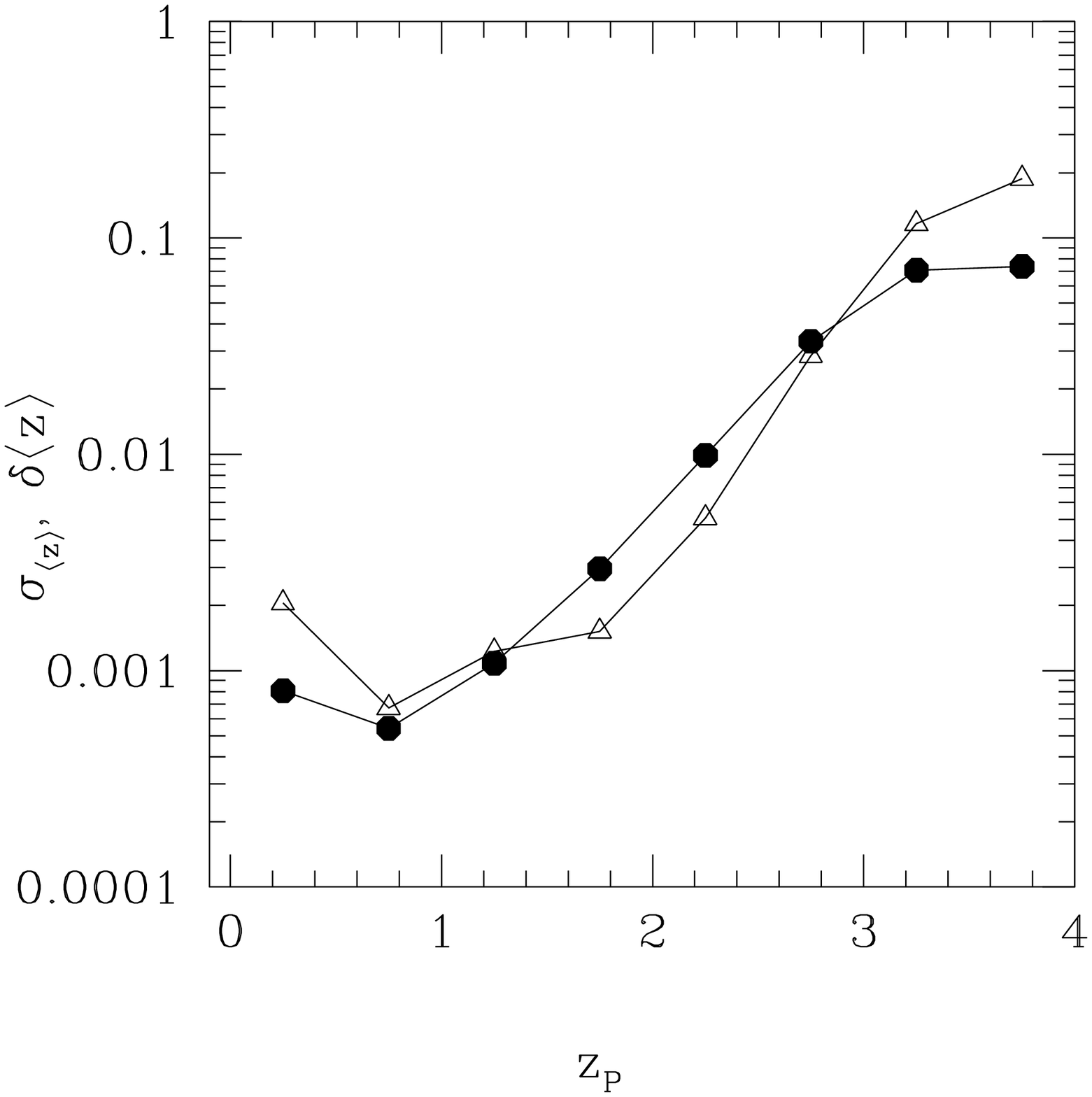}
\caption{The systematic error induced in the mean redshift  by the galaxy
  relative bias  
  $\tilde{b}$. Circles: statistical errors. Triangles: systematic
  errors. As a reminder, we adopt $\tilde{b}_{i\rightarrow
    j}=1+s(z_i-z_j)$ and $|s|=0.1$.  \label{fig:zmeanbias}} 
\efi

This galaxy distribution bias certainly biases
the $p_{i\rightarrow j}$ reconstruction. For example, if 
a subcategory of galaxies that did not cluster at all were to scatter, a cross
galaxy-galaxy correlation would not detect those, and result in the incorrect
leakage reconstruction. Interestingly, even for this extreme case of
  galaxy distribution bias, the galaxy-lensing correlation brings hope. For
  this subcategory of galaxies that scattered but did 
  not cluster  at all, other galaxies {\it apparently} behind of them may
  still be able to lens them and cause a detectable spurious foreground
  shear-background galaxy correlation. This example further demonstrates the
  gain by adding the galaxy-lensing cross correlation measurements in the
  self-calibration. 

Unfortunately, even with the aid of galaxy-lensing
correlation measurement, the self-calibration still fails, if no priors are
adopted. The galaxy distribution bias has not
only a deterministic component, but  also a stochastic component, i.e.
the noise induced by the different cosmic (sample) variance realizations
for different photo-z bins.
We show that, even if we can neglect the stochasticity, the 
degrees of freedom in the galaxy distribution bias kill the self-calibration. 

In this limit, the galaxy distribution bias can be completely described by
the relative bias parameter 
$\tilde{b}_{i\rightarrow j}$, namely the ratio of the bias between those 
scattered into 
the $j$-th redshift bin to those remaining in the $i$-th redshift bin, weighted
by the difference in $n(z)$.  With the  presence of the deterministic  galaxy
distribution  bias,  
Eq. \ref{eqn:gg} and Eq. \ref{eqn:Gg} become
\be
\label{eqn:ggb}
C^{gg,P}_{ij}\simeq \sum_{k}p_{k\rightarrow i}p_{k\rightarrow j}
\tilde{b}_{k\rightarrow i} \tilde{b}_{k\rightarrow j}C^{gg}_{kk}\ ,
\ee
\be
\label{eqn:Ggb}
C^{Gg,P}_{ij}=\sum_{k\geq m} p_{k\rightarrow i}p_{m\rightarrow j}
\tilde{b}_{m\rightarrow j}C^{Gg}_{km}\ .
\ee 
First of all, we notice a degeneracy in Eq. \ref{eqn:ggb}, of the form
$\tilde{b}_{i\rightarrow  j}\times p_{i\rightarrow j}$ ($i\neq j$). The same
argument helps to break the scaling  invariance of Eq. \ref{eqn:scaling} does
not apply here, simply due to many more free parameters involved here. The
galaxy-lensing correlation measurements do help, since the scaling 
invariance in Eq. \ref{eqn:ggb}, $p_{i\rightarrow j}\rightarrow
\tilde{b}_{i\rightarrow  j}\times p_{i\rightarrow j}$ ($i\neq j$),  does not
hold in Eq. \ref{eqn:Ggb}. Unfortunately, in general, $\tilde{b}(\ell)$ are 
scale 
dependent and there are $N_{\ell}N_z(N_z-1)$ of them, which nearly triple the
number of unknown parameters, making the number of unknowns larger than the
number of independent measurements and thus ruining the self-calibration.  

In reality, from the origins of the galaxy distribution
bias, we expect that it is scale dependent and is unlikely
deterministic. Thus, we are not able to render it as a statistical
error. Instead, we will live with it and quantify the induced bias in the 
  reconstructed $p$. 

\bfi{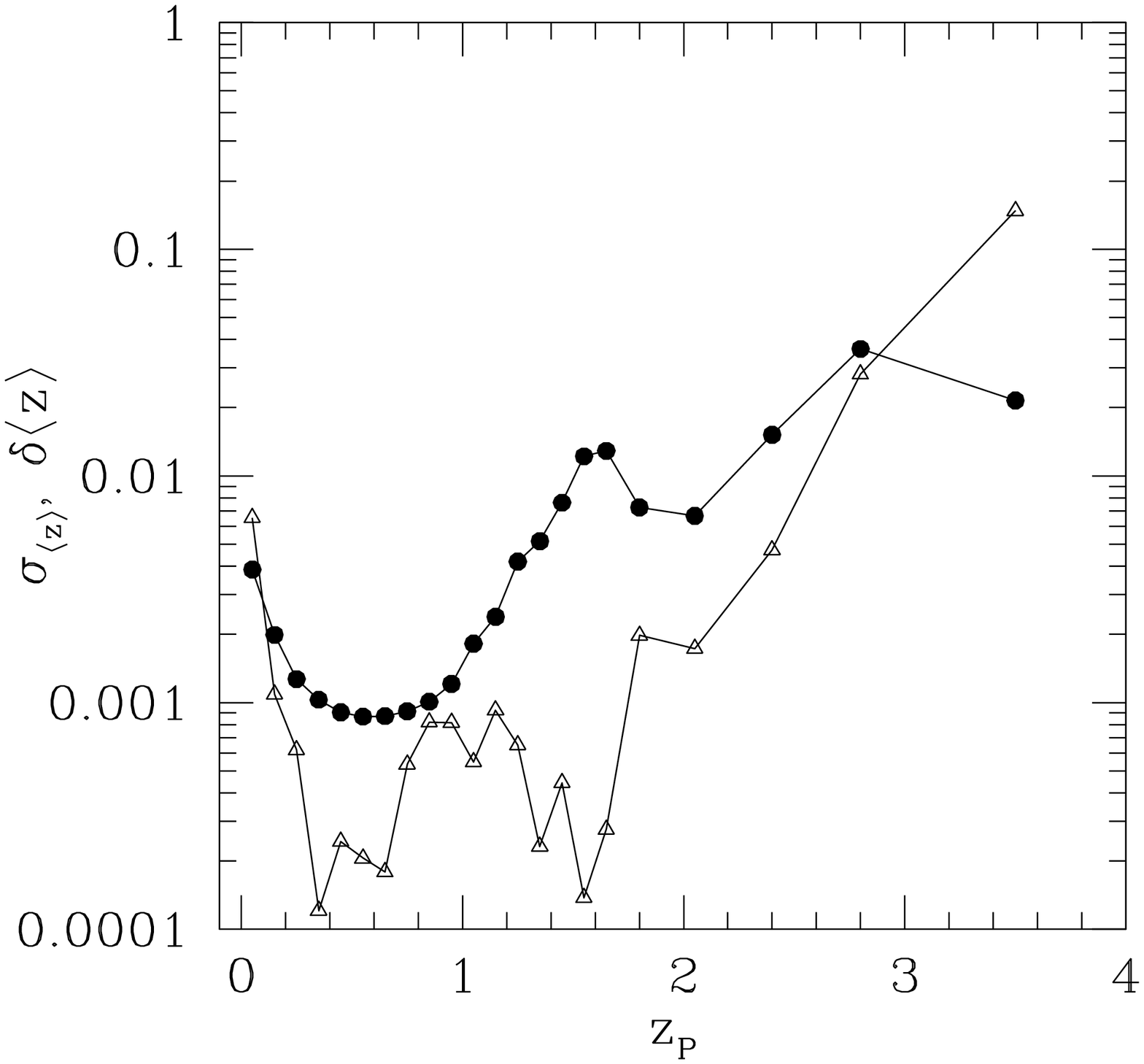}
\caption{The dependence of systematic error induced in the mean redshift  by
  the galaxy relative bias  
  $\tilde{b}$ on the size of redshift bins. Filled circles represent statistical
  errors and open triangles denote systematic 
  errors. As a reminder, we adopt $\tilde{b}_{i\rightarrow
    j}=1+s(z_i-z_j)$ and $|s|=0.1$.  \label{fig:zmeanbias22}} 
\efi

If we were to neglect this $\tilde{b}$ (namely by assuming $\tilde{b}=1$), the reconstructed $p$ would have a bias $\delta
p\sim (\tilde{b}-1)p$.   To robustly quantify the induced bias in $p$, we
need robust measurement or 
modeling of $\tilde{b}$, which we lack. To proceed,  we adopt a toy
model, $\tilde{b}_{i\rightarrow j}=1+s(z_i-z_j)$. The details of this
calculation are shown in the appendix. The resulting bias in $p$ scales as
$s$. For the case of $|s|=0.1$, the result is
shown in  Fig. \ref{fig:bias}, \ref{fig:zmeanbias} \& \ref{fig:zmeanbias22}. We
find that, 
for the most significant bias in 
$p$, it indeed satisfies the relation $\delta p\simeq (\tilde{b}-1)p$. For
those $p$ whose value is small, the dominant bias is induced by the propagation
from other parameters and thus do not follow this relation. 

Depending on the actual amplitude of this galaxy distribution bias, this may
be the dominant systematic error. It may also be non-negligible, or even
dominant, comparing to
the statistical errors in the reconstruction (Fig. \ref{fig:bias},
\ref{fig:zmeanbias} \& \ref{fig:zmeanbias22}). There are possible ways to 
reduce it. By choosing finer bin size, we can reduce the galaxy distribution
bias caused by the difference in $n(z)$, at the expense of more
$p_{i\rightarrow j}$ and $\tilde b_{i \rightarrow j}$ parameters to constrain.
With high-quality imaging or photometry we could further split galaxies into
sub-samples of morphological or spectral types to reduce the difference in
clustering strength.  If 
eventually we can reach $|s|<0.1$,  the galaxy distribution bias will not be
catastrophic, but still significant (Fig. \ref{fig:bias}).  

We caution that this galaxy distribution bias also exists in the calibration
technique based on cross correlations between photo-z and spec-z samples
\citep{Newman08,Bernstein09b}. In principle, direct spec-z sampling
of the photo-z galaxies \citep{Bernstein09b} allows for direct
measurement of  the galaxy distribution bias including its
stochasticity. Since the 
galaxy distribution bias has its own sample variance, the spec-z sampling must
be sufficiently wide in sky coverage, deep in redshift and reach high
completeness. These are crucial
issues for further investigation.

\subsection{The multiplicative error bias}
\label{sec:f}
Due to incomplete PSF correction, shear measurement can have multiplicative
errors and additive errors.The additive
errors do not bias the self-calibration results, since they do not
  correlate with galaxies. However, the multiplicative errors, which renders
  $\gamma$ to $(1+f)\gamma$,  can. If $f$ is the same for those galaxies whose
  photo-z remains in the true-z bin and those galaxies scatter out of the
  true-z bin, it does not induce bias in the $p$ reconstruction. However, in
  principle, these galaxies could have different multiplicative
  error. The multiplicative error could for example
  depend on the size of galaxies. If the photo-z error depends on some
  intrinsic properties of galaxies, which correlate with the galaxy size,
  then the multiplicative error would vary across different photo-z samples
  with the same true redshift. In this case, Eq. \ref{eqn:Gg} and
  \ref{eqn:Ggb} no longer hold. Eq. \ref{eqn:Ggb} should be replaced by 
\be
C^{Gg,P}_{ij}=\sum_{k\geq m} p_{k\rightarrow i}p_{m\rightarrow j}
(1+\Delta f_{k\rightarrow i}) \tilde{b}_{m\rightarrow
  j}\tilde{C}^{Gg,R}_{km}\ .
\ee
Here, the parameter $\Delta f_{k\rightarrow i}\equiv (1+f_{k\rightarrow
  i})/(1+f_{k\rightarrow k})-1\simeq f_{k\rightarrow i}-f_{k\rightarrow
  k}$ describes the relative difference in the multiplicative errors of the two
galaxy samples in the $k$-th true redshift bin (one scatters to 
the $i$-th photo-z bin and the other remains in the $k$-th photo-z
bin). $\tilde{C}^{Gg,R}_{km}=(1+f_{k\rightarrow k})C^{Gg,R}_{km}$. Clearly, if
$\Delta f=0$, it does not bias the $p$ reconstruction, since we just need to
redefine $C^{Gg,R}$. If $\Delta f\neq 0$, the
bias induced in $p$ is $\delta p \sim \Delta f p$. Current shape
  measurement  algorithms control $f$ to $\approx1\%$ 
levels with potentially larger redshift or size dependence \citep{STEP2}; but cosmic-shear analysis of future large surveys will
require $|f| \la 0.1\%$ if induced systematics are to be subdominant
to statistical  errors \citep{Huterer06, Amara08}.  Anticipating future
progress in shape measurement errors, we can expect an induced error $\la
0.001p$, which is sub-dominant to the one
induced by the galaxy distribution bias and is thus likely negligible. 
  However, in case that the shape measurement errors fail to reach the required
  accuracy, the bias induced by the relative  multiplicative error must be
  taken into account carefully.

\section{Dependence on the fiducial model}
\label{sec:fiducial}
The error forecast depends on the fiducial model we adopt, including the
  galaxy properties and the survey specifications. A thorough analysis over
  all uncertainties in the fiducial model is  beyond the scope of the current
  paper. Instead, we will present brief discussions on several key
  issues. 
\subsection{Dependence on the galaxy clustering properties}
In the above analysis, we have
made a number of simplifications. (1) We have adopted a scale independent and
redshift independent galaxy bias
for the error forecast. In reality, the galaxy bias is both redshift and scale
dependent. As we have seen, the reconstruction does not require assumptions on
the actual cosmology or clustering properties, so a variable bias is much
like changing the background cosmology, which in principle is independent
of the inferred scattering.
(2) All the power spectra in Eq. \ref{eqn:gg}, \ref{eqn:Gg}, etc.
are the ones in the observed cosmic volume. Due to the cosmic variance,
they can differ from the ensemble averages that we adopt for the fiducial
power spectra. As explained before, these uncertainties also affect the error
forecast. Eventually we will apply this self-calibration technique
to real data and thus will completely avoid the ambiguity in the fiducial
model. 

Here we will address the impact of scale dependent bias. As we 
have mentioned in \S \ref{sec:calibration}, the self-calibration with only
galaxy-galaxy clustering heavily relies on the shape differences between
different $C^{gg}_{ii}(\ell)$. If the 3D galaxy clustering is close to
power-law over a large scale range with redshift independent power index, the resulting
$C^{gg}_{ii}(\ell)$ are close to self-similar and thus the self-calibration with only galaxy-galaxy clustering will degrade significantly. The
question is, can the {\it full} self-calibration, with the aid from
galaxy-lensing cross correlation measurement, avoid this potential
degradation? 

\bfi{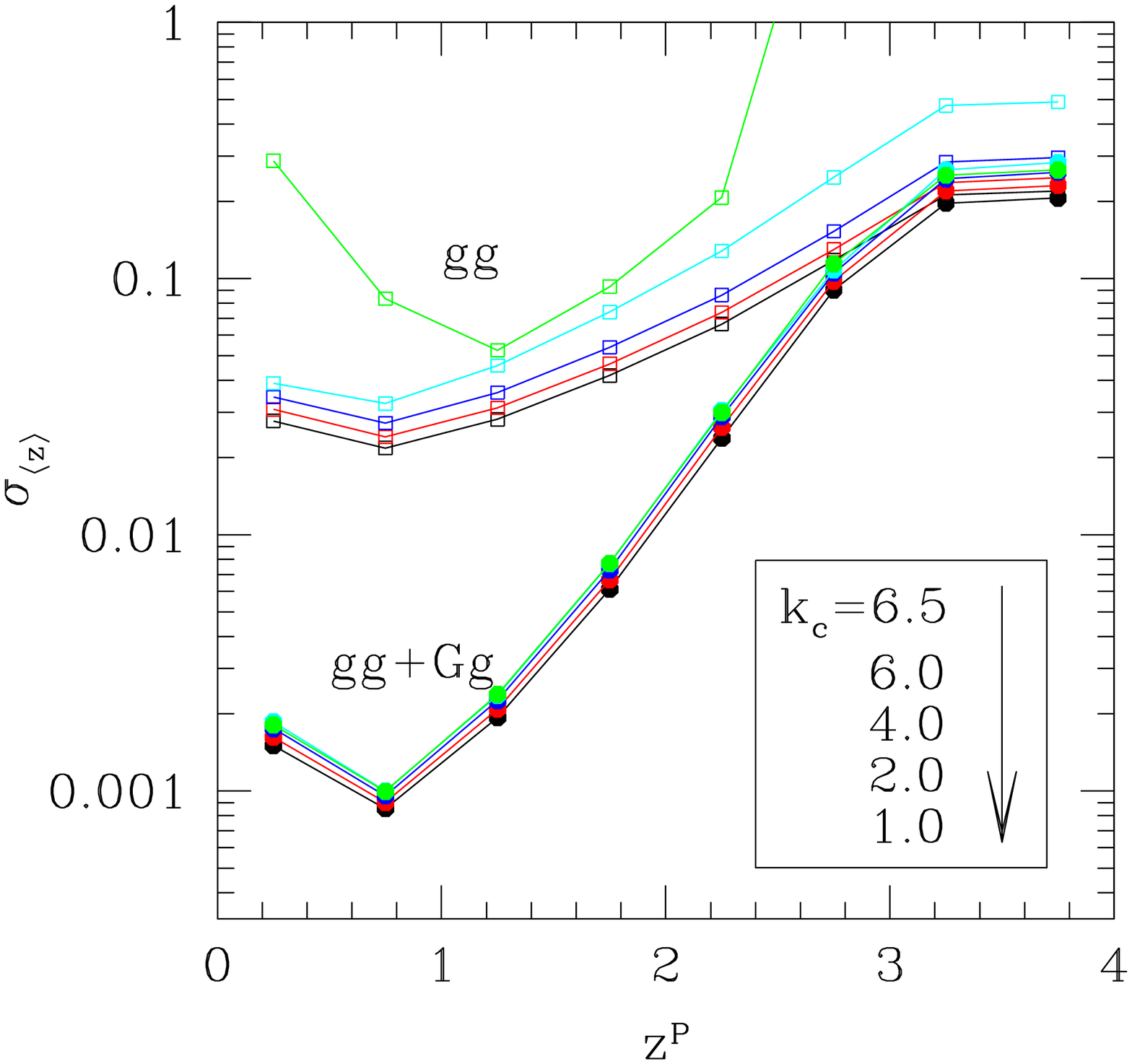}
\caption{The dependence of self-calibration performance on the shape of
  galaxy power spectrum. We adopt a toy model of galaxy clustering,
  characterized as a power law with a break at $k=k_c$. The shown value of
  $k_c$ is  in unit of $h/$Mpc. The ad-hoc
  galaxy power spectrum is closer to a power law for larger $k_c$. As
  expected, when the galaxy power spectrum is closer to a power law (with
  redshift-independent power index), the self-calibration based on galaxy
  clustering alone degrades and eventually blows up, due to more severe
  degeneracy between up and down photo-z scatters. On the other hand, the
  full self-calibration virtually avoids this problem, since it heavily relies
  on the intrinsic lensing geometry dependence to break this
  degeneracy.  \label{fig:powerlaw}} 
\efi 

To investigate this issue, we adopt an ad-hoc model for galaxy clustering,
in which the galaxy bias is scale dependent such that 3D galaxy power spectrum
(variance) takes the form
\be
\Delta_g^2(k,z)=10^{1.25}k^{1.85}(1+z)^{-2}\ \ when \ \ k<k_c\ ,
\ee  
and $\Delta_g^2(k,z)=\Delta_g^2(k_c,z)$ when $k\geq k_c$. If $k_c>1 h/$Mpc, at $z=0$ and
$0.1h/$Mpc $\la k<1 h/$Mpc, it is close to the matter power spectrum (variance)
$\Delta_m^2(k,z=0)$ and thus close to the galaxy clustering with $b_g=1$. But
it shows significant deviation from the matter power spectrum at other $k$ and
other redshifts. In the
limit that $k_c\rightarrow \infty$, this galaxy power spectrum becoms a strict
power-law and we expect that the self-calibration based on galaxy clustering
alone fails. Numerically, we find that when $k_c\ga 7h/$Mpc, the Fisher matrix
inversion based on galaxy clustering alone blows up, indicating its failure. 
 
Fig. \ref{fig:powerlaw} shows the degradation when increasing $k_c$. Despite
better galaxy clustering measurement due to stronger clustering strength, the
self-calibration accuracy degrades, since the galaxy power spectrum is closer
to a strict power-law  and the degeneracy between up and down photo-z scatters
becomes more severe. 

We expect that the full self-calibration is basically free of this problem,
since it mainly relies on the lensing geometry dependence to break the
degeneracy between up  and down scatters---namely, a lens can
only lens a  galaxy behind it. This is indeed what we find
numerically. To demonstrate this point, we assume the galaxy bias with respect
to the matter density to be deterministic, thus the 3D 
matter-galaxy cross correlation power spectrum (variance) is given by 
\be
\Delta^2_{mg}(k,z)=\sqrt{\Delta^2_g(k,z)\times \Delta^2_m(k,z)}\ .
\ee
This quantity determines the galaxy-galaxy lensing power spectrum. 
The accuracy of the full self-calibration is shown in
Fig. \ref{fig:powerlaw}.  The reconstruction accuracy only degrades slightly
even when the galaxy power spectrum is very close to a strict power-law
(e.g. $k_c=6.5h/$Mpc).  

This extreme example confirms our expectation that scale dependence of the
galaxy bias is unlikely to alter the major conclusions of this paper,
namely the  feasibility of our self-calibration technique.

\subsection{Dependence on other specifications}
The scaling of 
 $p_{i\rightarrow j}$ reconstruction accuracy with fiducial  quantities can be
    roughly understood as follows.   The observed density-density
correlations $C\propto p^2 b_g^2 \Delta^2_m$ and the shear-density correlations
$C\propto p^2 b_g \Delta^2_m$, where $\Delta^2_m$ is the matter power spectrum
(variance). These are the 
signals. For the noises in the correlation measurements, we have shown that shot  noises in shear and galaxy
number density measurements are the only relevant ones,  which scale as
 $\gamma_{\rm rms}(f_{\rm sky}\bar n_g^2)^{-1/2}$ and  $(f_{\rm sky}\bar n_g^2)^{-1/2}$
  respectively.  

Then if relying on galaxy-galaxy
lensing measurement alone, the reconstruction  error $\sigma_p\propto
\gamma_{\rm rms}^{1}$. In combination with galaxy-galaxy 
    clustering measurement, the dependence becomes weaker and we expect that
$\sigma_p\propto \gamma_{\rm rms}^{e}$, where $e\in (0,1)$. If we
    adopt a fiducial value $\gamma_{\rm rms}=0.24$ instead of $0.2$, the
    reconstruction accuracy will degrade by a factor of $<20\%$. 

For fixed galaxy distribution, matter power spectrum $\Delta^2_m$, and $p_{i\rightarrow
  j}$, following similar argument above, we find that
\be
\sigma_p\propto f_{\rm  sky}^{-1/2}\bar{n}_g^{-1}b_g^{-d}\gamma_{\rm rms}^{e}\ .
\ee 
Here, the bias dependence $d\in (1,2)$ and, as a reminder, $e\in (0,1)$. 

 It is interesting to quantify the performance of the self-calibration
  technique for surveys like CFHTLS and DES.  Based on
  the above scalings, we are able to do an order 
  of magnitude estimation.  (1) One of the major differences between these
  surveys and the fiducial stage 
IV survey, is the sky coverage. From the $f_{\rm sky}$ dependence alone, for
CFHTLS, we expect a factor of 10 larger reconstruction 
errors, since the sky coverage is a factor of 100 smaller. For DES, we expect
the degradation to be a factor of 2. Since the statistical accuracy of these
surveys scales exactly the same way with respect to the sky coverage $f_{\rm
  sky}$, the self-calibration technique works equally fine for these surveys,
from this viewpoint. (2) Another difference is that the number
density of source galaxies in CFHTLS and DES is likely a factor of 2
smaller. This results in another factor of 2 degradation in the reconstruction
accuracy.  Since the lensing measurement at $\ell<2000$ is not completely shot
noise dominated, the statistical accuracy has weaker dependence on
$\bar{n}_g$. From this viewpoint, the self-calibration technique works better
for surveys with higher galaxy number density.

We caution that the above estimation neglects many
complexities. For example, CFHTLS, DES and Pan-STARRS are shallower than the
fiducial survey and thus the galaxy number densities at high redshift in these
surveys are likely much smaller, while the galaxy number densities at low
redshift are comparable. This implies
that the $p_{i\rightarrow j}$ reconstruction in these surveys at high-z is
more affected that at low-z. Furthermore, the above estimation neglects the
difference in the  photo-z error, which is likely considerably worse for
CFHTLS and DES. It will definitely affect the reconstruction. However, as we
discussed  in \S\ref{subsec:error}, the detection threshold of contamination rate
($p_{i\rightarrow j}$) is mainly determined by the ratio of noise and signal
of corresponding true-z bins. In this sense,  the self-calibration technique
works better for worse photo-z estimation.

For a robust forecast of the performance of our self-calibration technique in
each specific survey, we need a detail fiducial model of the galaxy 
distribution, clustering  and photo-z error distribution.
Although this task is beyond the scope of  this paper, the above
estimates imply that the self-calibration technique will be applicable.

\section{Discussions}
\label{sec:discussion}
It is possible to further improve the statistical accuracy of the
self-calibration technique. For example, so far we treat $C^{Gg}_{i>j}$ and
$C^{Gg}_{k>j}$ as independent quantities. Improvement can be made by utilizing
their internal connection. Both of them are determined
by the same mass-galaxy cross correlation over the same redshift range. For
this reason, they are connected by a simple scaling relation (in the absence of 
intrinsic alignments), as pointed out
by \citet{Jain03,Zhang05,Bernstein06}. 

In weak lensing
cosmology, people often disregard the lensing power spectrum measurement  at
$\ell>2000$-$3000$, because theoretical prediction at such scale is largely
uncertain. However, the shear-shear measurement at such scales contains useful
information to improve the photo-z reconstruction as well as shear-ratio information that is useful for cosmology. For example, they allow for
better handle over the magnification and size bias induced correlations,
mainly in the background density-foreground shear correlation measurement.
We could use this information usually disregarded in cosmological applications
to improve  the photo-z calibration. 

So far we only focus on the reconstructed $p_{i\rightarrow j}$. The
reconstructed $C^{gg,R}_{ii}$ and $C^{Gg,R}_{i\geq j}$ contain valuable
information on cosmology and can be further explored. Since these
reconstructed power spectra do not suffer from the problem of the photo-z
scatters, they can be utilized for purpose beyond cosmology. For example, the self-calibration of galaxy intrinsic alignment, proposed by
\citet{Zhang08}, relies on the measurement of $C^{Gg,P}_{ii}$ to infer the GI
correlation contaminated the weak lensing measurement.  This quantity is
contaminated by the photo-z scatters.  Since we are now able to quantify the
photo-z scatters and $C^{Gg,R}_{ij}$ simultaneously, we are able to quantify
and correct for the effect of photo-z scatters in the self-calibration of
galaxy intrinsic alignment.

One key issue missing in this paper is to propagate the errors or biases on
the reconstructed $p_{i\rightarrow j}$ into errors on cosmology inferred from
shear-shear data. This will tell us whether the errors in the self-calibration
 are small enough to avoid significant biases or inflated errors in a
 shear-shear cosmology measurement.  Although many of the $p$'s are uncertain by
 more  than the $\sim 0.1\%$ that is needed to make the
 biases  negligible \citep{Bernstein09b},  there are many correlations between
 these errors which complicate the estimation. This issue definitely deserves
 further investigation.

We emphasize that our purely photometric self-calibration technique is
complementary 
  to those based on cross correlations between photo-z and spec-z samples
  \citep{Newman08,Bernstein09}, which we call cross-calibration. An
  advantage of cross-calibration is that it can identify a special type of
  photo-z error, namely when the mean photo-z is a monotonic increasing
  function of 
  true-z (other than the correct identity function).  As explained earlier, the self-calibration technique fails
  completely for such type of photo-z error.  The cross-calibration technique will also likely be more able to infer galaxy distribution bias caused when photo-z outliers have a different $n(z)$ than other galaxies in the same true-z bin.
On the other hand, the 
  self-calibration has a number of advantage over the cross correlation. Since
the total number of photo-z galaxies is much larger than that of the spec-z
sample, it can reach higher statistical accuracy. Since what it measures is
the photo-z scatters in the whole survey volume, it avoids possible cosmic
variance in the photo-z scatters, which could bias the cross-calibration. And
since the spec-z targets may be a very different population from the photo-z
galaxies, the cross-correlation method will be more susceptible to biases from
varying $b_g$ among subpopulations. 

The photo-z scatter self-calibration method described here has many attractive
aspects: it does not depend on any cosmological priors or on models of the
power spectrum; it is unaffected by intrinsic alignments; its errors are
determined by shot noise, not sample variance, so that higher source densities
are exploited if observed; it remains Gaussian and tractable to high $\ell$;
and is not significantly affected by shear measurement errors.  And of course
it can be conducted with the same imaging data used for the shear-shear
correlation measurement without degrading the shear-shear information content.
The Fisher analysis suggests that photo-z outlier rates can be determined with
statistical errors of 0.01--1\% for bins at $z\le2$.  It will be necessary to
correct data for lensing magnification bias, and in principle this can be done
with little statistical penalty, but the magnification bias factor $g$ must be
determined to sufficient accuracy.  The biggest issue is ``galaxy distribution
bias,'' whereby photo-z outlier galaxies might have different $n(z)$ or bias
$b_g$ than the non-outlying galaxies in the same true-z bin.  We find that the
effective bias of galaxies must vary by $<O(0.1)$ in order to avoid systematic
errors in scattering rates that exceed the expected statistical errors.  This
is an area deserving of more detailed attention. 

In this paper, we have
  presented a concept study of the proposed self-calibration. For the
  idealized survey of stage IV survey specifications, we have
  shown that it can in principle reconstruct the photo-z error distribution to
  useful precision. More robust forecast shall 
include all extra sources of error, as listed in \S \ref{sec:statistical} and
\S \ref{sec:bias}, more realistic fiducial model (\S \ref{sec:fiducial}), and
possibly more uncertainties,  into account. 

\section{Acknowledgment}
We thank Hu Zhan, Sarah Bridle and Jun Pan for many
useful discussions and the anonymous referee for many useful suggestions.  PJZ 
thanks the hospitality of the UPenn physics and astronomy department and the
Aspen center for physics, where part of the work was done. 
PJZ thanks the support of the one-hundred talents program of the Chinese
academy of science, the national science
foundation of China (grant No. 10533030, 10543004, 10821302 \& 10973027), the
CAS grant  
KJCX3-SYW-N2 and the 973 program grant No. 2007CB815401.
GMB acknowledges support from grant AST-0607667 from the
National Science Foundation and Department of Energy grant
DOE-DE-FG02-95ER40893.

\onecolumn
\appendix
\section[]{The likelihood analysis and the Fisher matrix layout}
\label{sec:appendixA}
We work on the likelihood of the band angular power spectra to quantify the self-calibration error. As explained in \S \ref{sec:calibration}, 
the error estimation here is distinctly different from that
in the routine  exercises of cosmological parameter constraints and  the only
relevant noise term is the 
shot noise. This makes the error estimation much simpler. Here we recast this
argument in a more formal way. 

The data we have are a set of measured power spectra contaminated by shot
noise, $C^{D}=C^{P}+\delta C^S$. Here, the superscript ``D'' denotes the
data. $\delta C^S$ is the fluctuation in shot noise.
$C^P$ are the power spectra in
photo-z bins, which are related to $C^R$ by a set of
$p_{i\rightarrow j}$ through Eq. \ref{eqn:gg} \& \ref{eqn:Gg}.
We want to know the likelihood function $P(p_{i\rightarrow
  j}|C^{D})$. Since the distribution of $C^D$ is completely determined by
$p_{i\rightarrow j}$ and $C^R$ (given the shot noise distribution), 
Bayes' theorem states
\be
P(p_{i\rightarrow
  j},C^R|C^{D})\propto P(C^{D}|p_{i\rightarrow
  j},C^R)P(p_{i\rightarrow
  j},C^R)=P(C^{D}|p_{i\rightarrow
  j},C^R)P(p_{i\rightarrow j})P(C^R) \ .
\ee
We then have
\be
\label{eqn:pij}
P(p_{i\rightarrow j}|C^{D})=\int P(p_{i\rightarrow
  j},C^R|C^{D})dC^R\propto  \int P(C^{D}|p_{i\rightarrow
  j},C^R)P(p_{i\rightarrow j})P(C^R)dC^R\ .
\ee 
For $p_{i\rightarrow j}$, we only exert the condition $\sum_i p_{i\rightarrow
  j}=1$. Since we do not take any cosmological prior, we set the prior
$P(C^R)=1$. Then  Eq. \ref{eqn:pij} means to marginalize over all possible
value of $C^R$. This 
is what we have done in the paper. Notice that $P(C^{D}|p_{i\rightarrow
  j},C^R)$ is completely determined by the shot noise
distribution, given $C^R$ and $p_{i\rightarrow j}$, the parameters that we
want to fit. That is the reason that we state that only shot noise shows up in
the error analysis. And, since we do not have real data, we
have to arbitrarily choose $C^{D}$, which could differ from the cosmic mean,
due to the existence of cosmic variance. This freedom is where the cosmic
variance shows up. In the paper, we follow the common choice and set $C^{D}=\langle
C^R\rangle$, the ensemble 
average. But this is just a representative case of the real data. Namely,
$C^{D}$ can be any other value in the range allowed by the cosmic variance and
the shot noise, as discussed in one of the footnotes in \S \ref{subsec:error}. 

In Eq. \ref{eqn:pij}, we can take stronger prior on $C^R$. In principle, if we
know the cosmology, 
we can write down the full PDF of $C^R$.  In reality, we do not know exactly
the cosmology and we do not know exactly the PDF in the non-linear,
non-Gaussian regime. Nevertheless, these priors will improve the
reconstruction accuracy of $p_{i\rightarrow j}$. However, as we show in the
paper, the systematical error induced by galaxy distribution bias may have
already been the dominant error source, improvement over statistical error is
not necessary at the current moment.

Since the fluctuations are induced by the shot noise, the central limit
theorem implies that we can approximate the probability distribution of the
power spectra $P(C^D|p_{i\rightarrow j},C^R)$ as Gaussian, as long as there are sufficiently large number of independent
$\vec{\ell}$  modes in each multipole $\ell$ bin ($\ell\Delta {\ell}f_{\rm
  sky}\gg 1$). Since we  only use those 
modes with $\ell>100$ and $\Delta l\gg 1$, the probability distribution of the
power spectra should be Gaussian for $f_{\rm sky}\ga 0.01$. Furthermore, the
covariance 
matrix is diagonal, namely, different power spectrum 
measurements are uncorrelated, since the associated shot noises are
uncorrelated. The covariance matrix between the power spectra
$C^{(\alpha),P}_{i_1j_1}$ and $C^{(\beta),P}_{i_2j_2}$, with 
$(\alpha),(\beta)=gg,Gg$,  is 
\be
C^{(\alpha),(\beta)}_{i_1j_1;i_2j_2}=\left[\sigma_{i_1j_1}^{(\alpha)}\right]^2\delta_{(\alpha)(\beta)}\delta_{i_1i_2}\delta_{j_1j_2}\ ,
\ee
with 
\be
(\sigma^{gg}_{ij})^2=\frac{\left[4\pi f_{\rm sky}/N_i\right]\left[4\pi
      f_{\rm sky}/N_j\right]}{2\ell\Delta \ell f_{\rm
    sky}}(1+\delta_{ij})\ ,\ (\sigma^{Gg}_{ij})^2=\frac{\left[4\pi f_{\rm sky} \gamma^2_{\rm
        rms}/N_i\right]\left[4\pi 
      f_{\rm sky}/N_j\right]}{2\ell\Delta \ell f_{\rm sky}}\ .
\ee

 The logarithm of the likelihood function is
\be
\label{eqn:likelihood}
\ln L=-\frac{1}{2}\sum_{\ell} \left[\sum_{i\leq j}
\left(\sigma^{gg}_{ij}(\ell)\right)^{-2}\left(C^{gg,D}_{ij}(\ell)-C^{gg,P}_{ij}(\ell)\right)^2
+ \sum  \left(\sigma^{Gg}_{ij}(\ell)\right)^{-2}
\left(C^{Gg,D}_{ij}(\ell)-C^{Gg,P}_{ij}(\ell)\right)^2\right]+{\rm const}. \ \ .
\ee
Here, $C^{gg,P}_{ij}$ and $C^{Gg,P}_{ij}$ are the model prediction, given by
Eq. \ref{eqn:gg} \& \ref{eqn:Gg}. If the model is unbiased, 
we can set $C^{gg,D}=C^{gg,P}$ and $C^{Gg,D}=C^{Gg,P}$ when evaluating the Fisher matrix $F_{ij}\equiv \partial^2 \ln L/\partial \lambda_i\partial
 \lambda_j$. We then have 
\ba
\label{eqn:F}
{\bf F}_{km}=\sum_{\ell}\left[\sum_{i\leq j}
\left(\sigma^{gg}_{ij}(\ell)\right)^{-2} \frac{\partial
  C^{gg,P}_{ij}(\ell)}{\partial \lambda_k} \frac{\partial
  C^{gg,P}_{ij}(\ell)}{\partial \lambda_m}
+ \sum_{ij}  \left(\sigma^{Gg}_{ij}(\ell)\right)^{-2} \frac{\partial
  C^{Gg,P}_{ij}(\ell)}{\partial \lambda_k} \frac{\partial
  C^{Gg,P}_{ij}(\ell)}{\partial \lambda_m}\right]\ .
\ea
$C^{gg,P}_{ij}$ and $C^{gg,P}_{ji}$ are equivalent, so we only need to
sum up pairs with $i\leq j$ for the galaxy-galaxy correlation. On the other
hand, for the lensing-galaxy cross correlation, we should sum up all
pairs (in this case, $ij$ and $ji$ pairs are asymmetric).  The unknown parameters to be determined are $\lambda=(p_{\mu\rightarrow
  \nu},C^{Gg,R}_{ij}(\ell_1), C^{gg,R}_{kk}(\ell_1)\ldots)$, with $\mu\neq
\nu$, $i\geq j$ and $\mu,\nu,i,j,k=1,\ldots$. Here, $\ell_{1,2\ldots}$ denote
the multipole $\ell$ bins.  The Fisher matrix can be decomposed as follows,
\be
\label{eqn:FL}
{\bf F}=\bordermatrix{ & & \cr
                &{\bf A}&{\bf B}\cr
                &{\bf C}&{\bf D}\cr}=
\bordermatrix{
& & & &\cr
&\sum_{\ell} {\bf A_{\ell}}&{\bf B}_{\ell_1}&\ldots&{\bf B}_{\ell_n}\cr
&{\bf C}_{\ell_1}         &{\bf D}_{\ell_1}&\ldots         &0 \cr
&\vdots                 & \vdots        &\ddots         &\vdots\cr
& {\bf C}_{\ell_n}       & 0              &\ldots         &{\bf
    D}_{\ell_n}\cr}\ .
\ee
Here, ${\bf A}\equiv {\bf F}_{p_{\mu_1\nu_1}p_{\mu_2\nu_2}}$. It is the sum of
${\bf A}_{\ell}$, the contribution from each $\ell$ 
bin.   ${\bf B}\equiv
{\bf F}_{p_{\mu_1\nu_1}C_{\mu_2\nu_2}}$ and  ${\bf C}={\bf B}^T$. ${\bf
  B}$ can be decomposed as ${\bf B}=({\bf B}_{\ell_1},{\bf
  B}_{\ell_2},\ldots)$, where ${\bf B}_{\ell}$ is the contribution from each
$\ell$ bin. ${\bf
  D}={\bf F}_{C_{\mu_1\nu_1}C_{\mu_2\nu_2}}$.  Since different $\ell$ bins do
not correlate, ${\bf D}$ is block diagonal and denote each block element as
${\bf D}_{\ell}$. The exact definitions of ${\bf A}_{\ell}$, ${\bf B}_{\ell}$,
${\bf C}_{\ell}$ and ${\bf D}_{\ell}$ can be found by comparing
Eq. \ref{eqn:likelihood} and \ref{eqn:FL}. The dimension of ${\bf A}$ and
${\bf A}_l$ is $N_z(N_z-1)\times N_z(N_z-1)$. That of ${\bf B}_{\ell}$ is
$N_z(N_z-1)\times N_z(N_z+3)/2$ and that of ${\bf D}_{\ell}$ is
N$_z(N_z+3)/2\times N_z(N_z+3)/2$.    The inverse of ${\bf F}$ is
\be
\label{eqn:IF}
{\bf F}^{-1}=\bordermatrix{ & & \cr
                &\left({\bf A}-{\bf B}{\bf D}^{-1}{\bf
    C}\right)^{-1}&-\left({\bf A}-{\bf B}{\bf D}^{-1}{\bf C}\right)^{-1} {\bf
    B}{\bf D}^{-1}\cr
                &-{\bf D}^{-1}{\bf C}\left({\bf A}-{\bf B}{\bf D}^{-1}{\bf
    C}\right)^{-1}&{\bf D}^{-1}+{\bf D}^{-1}{\bf C}\left({\bf A}-{\bf B}{\bf D}^{-1}{\bf
    C}\right)^{-1}{\bf B}{\bf D}^{-1}\cr}\ .
\ee
For convenience, we define
\be
\label{eqn:reducedA}
{\bf A}^{\rm reduced}\equiv {\bf A}-{\bf B}{\bf D}^{-1}{\bf
  C}=\sum_{\ell}\left[{\bf A}_{\ell}-{\bf B}_{\ell}{\bf D}^{-1}_{\ell}{\bf
    C}_{\ell}\right]\equiv \sum_{\ell} {\bf A}^{\rm reduced}_{\ell}\ .
\ee
The second expression holds because ${\bf D}$ is block diagonal.  The reduced
matrix ${\bf A}^{\rm reduced}$ determines the reconstruction accuracy of
$p_{i\rightarrow j}$. The last expression simply means that, since $\ell$ bins
are uncorrelated, we are able to sum over the contribution from each of them
and improve the reconstruction accuracy. We know that constrains on the
photo-z scatters mainly come from $C^{Gg,P}_{i<j}$ and $C^{gg,P}_{i\neq
  j}$. So we can further
decompose ${\bf A}_{\ell}={\bf A}^{\rm min}_{\ell}+\Delta {\bf A}_{\ell}$,
where ${\bf A}^{\rm min}_{\ell}$ is the contribution from $C^{Gg,P}_{i<j}$ and $C^{gg,P}_{i\neq
  j}$ and $\Delta {\bf A}_{\ell}$ is the contribution from $C^{Gg,P}_{i\geq j}$
and $C^{gg,P}_{i=j}$. For clarity, we provide the expression of $\Delta {\bf
  A}_{\ell}$, 
\ba
\label{eqn:A}
\Delta {\bf A}_{\ell,km}=\sum_{i=j}
\left(\sigma^{gg}_{ij}(\ell)\right)^{-2} \frac{\partial
  C^{gg,P}_{ij}(\ell)}{\partial \lambda_k} \frac{\partial
  C^{gg,P}_{ij}(\ell)}{\partial \lambda_m} 
+ \sum_{i\geq j}  \left(\sigma^{Gg}_{ij}(\ell)\right)^{-2} \frac{\partial
  C^{Gg,P}_{ij}(\ell)}{\partial \lambda_k} \frac{\partial
  C^{Gg,P}_{ij}(\ell)}{\partial \lambda_m}\ .
\ea
Here, $\lambda_k,\lambda_m\in p_{\mu\nu}$.

In the limit $p_{\mu \neq \nu}\rightarrow 0$ (equivalently $p_{\mu\rightarrow
  \nu}\rightarrow \delta_{\mu\nu}$), we have $\Delta {\bf A}_{\ell}\rightarrow
{\bf 
  B}_{\ell}{\bf D}^{-1}_{\ell}{\bf C}_{\ell}$. The proof is as follows. since
$p_{i\rightarrow j}=\delta_{ij}$, ${\bf D}_{\ell}$
is diagonal, with  the diagonal elements $
{\bf 
  D}_{C^{Gg}_{\mu_1\nu_1};C^{Gg}_{\mu_1\nu_1}}=(\sigma^{Gg}_{\mu_1\nu_1})^{-2}$
and $
{\bf
  D}_{C^{gg}_{\mu_1\mu_1};C^{gg}_{\mu_1\mu_1}}=(\sigma^{gg}_{\mu_1\mu_1})^{-2}$. 
We then have
\ba
{\bf
  B}_{\ell}{\bf D}^{-1}_{\ell}{\bf C}_{\ell}=\sum_{i=j}
\left(\sigma^{gg}_{ij}(\ell)\right)^{-2} \frac{\partial
  C^{gg,P}_{ij}(\ell)}{\partial \lambda_k} \frac{\partial
  C^{gg,P}_{ij}(\ell)}{\partial \lambda_m} 
+\sum_{i\geq j}  \left(\sigma^{Gg}_{ij}(\ell)\right)^{-2} \frac{\partial
  C^{Gg,P}_{ij}(\ell)}{\partial \lambda_k} \frac{\partial
  C^{Gg,P}_{ij}(\ell)}{\partial \lambda_m}= \Delta {\bf A}_{\ell}\ .
\ea
Thus under the limit that $p_{i\neq j}\rightarrow 0$, ${\bf A}^{\rm
  reduced}={\bf A}^{\rm min}$ and the the error matrix of $p$ is
$[{\bf A}^{\rm min}]^{-1}$. This result has a clear physical meaning. We know
that the measurements $C^{gg,P}_{i=j}$ and $C^{Gg,P}_{i\geq j}$ are most
responsible for determining $C^{gg,R}_{i=j}$ and $C^{Gg,R}_{i\geq j}$ and  the
measurements $C^{gg,P}_{i\neq j}$ and $C^{Gg,P}_{i<j}$ are most responsible to
determine $p$. Under the limit $p_{i\neq j}\rightarrow 0$, we are able to have
a clear separation of the two sets of measurements. 

We caution that  with the presence of photo-z scatters, this is no longer the
case. Numerically, we find that the presence 
of photo-z scatters make the statistical error in $p$ reconstruction, larger
than the above limiting case.  

We now derive the detailed expression of all the matrices. A useful relation in
this exercise is that $\partial p_{k\rightarrow i}/\partial
p_{\mu\rightarrow \nu}=\delta_{i\nu}(\delta_{k\mu}-\delta_{k\nu})$. The last
term shows up because $p_{i\rightarrow i}=1-\sum_{k\neq i}p_{k\rightarrow i}$
and only $p_{\mu\neq \nu}$ are independent variables. We then have 
\ba
\frac{\partial C_{ij}^{(\alpha),P}}{\partial p_{\mu\rightarrow
      \nu}}=\delta_{i\nu}\sum_{m}\left(C_{\mu m}^{(\alpha),R}-C_{\nu
      m}^{(\alpha),R}\right)p_{m\rightarrow
      j}+\delta_{j\nu}\sum_{k}\left(C_{k\mu}^{(\alpha),R}-C_{k\nu}^{(\alpha),R}\right)p_{k\rightarrow
    i}\equiv \delta_{i\nu} A_{\mu\nu j}^{(\alpha)}+\delta_{j\nu}B_{\mu\nu
  i}^{(\alpha)}\ .
\ea
Clearly, we have $A_{\mu\nu i}^{gg}=B_{\mu\nu i}^{gg}$. But there is no such
relation for the G-g terms.  Finally we have 
\ba
{\bf A}^{\rm min}_{\ell,p_{\mu_1\rightarrow \nu_1},p_{\mu_2\rightarrow
    \nu_2}}&=&
\sum^{(\alpha)}\left[\left(\sigma_{\nu_1\nu_2}^{(\alpha)}\right)^{-2}A_{\mu_1\nu_1\nu_2}^{(\alpha)}B_{\mu_2\nu_2\nu_1}^{(\alpha)}H(\nu_2-\nu_1)
+\left(\sigma_{\nu_2\nu_1}^{(\alpha)}\right)^{-2}B_{\mu_1\nu_1\nu_2}^{(\alpha)}A_{\mu_2\nu_2\nu_1}^{(\alpha)} H(\nu_1-\nu_2)\right.
  \nonumber \\
&+&\left. \sum_{j>\nu_1} \left(\sigma_{\nu_1 j}^{(\alpha)}\right)^{-2}A_{\mu_1\nu_1 j}^{(\alpha)}A_{\mu_2\nu_2
  j}^{(\alpha)}\delta_{\nu_1\nu_2} + \sum_{i<\nu_1}
  \left(\sigma_{i\nu_1}^{(\alpha)}\right)^{-2}B_{\mu_1\nu_1i}^{(\alpha)}B_{\mu_2\nu_2i}^{(\alpha)}\delta_{\nu_1\nu_2}
  \right] \ ,
\ea
\ba
\Delta {\bf A}_{\ell,p_{\mu_1\rightarrow \nu_1},p_{\mu_2\rightarrow
    \nu_2}}&=&\delta_{\nu_1\nu_2}\left(\sigma_{\nu_1\nu_1}^{(gg)}\right)^{-2}\times(A_{\mu_1\nu_1\nu_1}^{(gg)}+B_{\mu_1\nu_1\nu_1}^{(gg)})(A_{\mu_2\nu_1\nu_1}^{(gg)}+B_{\mu_2\nu_1\nu_1}^{(gg)})\nonumber\\
&+&\left(\sigma_{\nu_1\nu_2}^{(Gg)}\right)^{-2}A_{\mu_1\nu_1\nu_2}^{(Gg)}B_{\mu_2\nu_2\nu_1}^{(Gg)}H(\nu_1-\nu_2) +\left(\sigma_{\nu_2\nu_1}^{(Gg)}\right)^{-2}B_{\mu_1\nu_1\nu_2}^{(Gg)}A_{\mu_2\nu_2\nu_1}^{(Gg)} H(\nu_2-\nu_1)
  \nonumber \\
&+&\sum_{j\leq \nu_1} \left(\sigma_{\nu_1 j}^{(Gg)}\right)^{-2}A_{\mu_1\nu_1 j}^{(Gg)}A_{\mu_2\nu_2
  j}^{(Gg)}\delta_{\nu_1\nu_2} +\sum_{i\geq \nu_1}
  \left(\sigma_{i\nu_1}^{(Gg)}\right)^{-2}B_{\mu_1\nu_1i}^{(Gg)}B_{\mu_2\nu_2i}^{(Gg)}\delta_{\nu_1\nu_2}
  \ ,
\ea
\ba
{\bf B}_{\ell,p_{\mu_1\rightarrow \nu_1},C^{Gg}_{\mu_2\nu_2}}=\sum (\sigma_{\nu_1j}^{Gg})^{-2}A_{\mu_1\nu_1j}p_{\mu_2\rightarrow\nu_1}p_{\nu_2\rightarrow
  j}+\sum_{i\neq
  \nu_1}
(\sigma_{i\nu_1}^{Gg})^{-2}B_{\mu_1\nu_1i}p_{\mu_2\rightarrow
  i}p_{\nu_2\rightarrow\nu_1} \ ,
\ea
\ba
{\bf B}_{\ell,p_{\mu_1\rightarrow \nu_1},C^{gg}_{\mu_2\nu_2}}=\sum_{j\geq
  \nu_1}(\sigma_{\nu_1j}^{gg})^{-2}A_{\mu_1\nu_1j}p_{\mu_2\rightarrow\nu_1}p_{\nu_2\rightarrow
  j}+\sum
(\sigma_{i\nu_1}^{gg})^{-2}B_{\mu_1\nu_1i}p_{\mu_2\rightarrow
  i}p_{\nu_2\rightarrow\nu_1}\ ,
\ea
\ba
{\bf D}_{\ell,C^{Gg}_{\mu_1\nu_1},C^{Gg}_{\mu_2\nu_2}}=\sum
(\sigma_{ij}^{Gg})^{-2}p_{\mu_1 \rightarrow i}p_{\nu_1 \rightarrow j}p_{\mu_2
  \rightarrow i}p_{\nu_2 \rightarrow j}\ ,
\ea
and 
\ba
{\bf D}_{\ell,C^{gg}_{\mu_1\nu_1},C^{gg}_{\mu_2\nu_2}}=\sum_{i\leq j}
(\sigma_{ij}^{gg})^{-2}p_{\mu_1 \rightarrow i}p_{\nu_1 \rightarrow j}p_{\mu_2
  \rightarrow i}p_{\nu_2 \rightarrow j}\ ,
\ea
Here, $H(x)$ is the Heaviside step function, whose value is $1$ when $x$ is
positive, otherwise zero. In the above equations, all the power spectra at the
right hand sides are evaluated at the corresponding $\ell$. 

We now only need to invert the matrix ${\bf D}_{\ell}$ and ${\bf A}^{\rm
  reduced}$. This is a much easier job than the inversion of the full Fisher
matrix 
${\bf F}$. More than that, it improves the numerical stability and accuracy of
the inversion, which we can check at each step according to Eq. \ref{eqn:IF} \&
\ref{eqn:reducedA}.  Numerically, we do not find any numerical instability or
singularity for all the configurations that we have tested. The reconstruction
accuracy of $p$ is solely determined by ${\bf A}^{\rm
  reduced}$ and the results are shown in Fig. \ref{fig:p8}. 

Alternative to this detailed description of the photo-z errors, we can
evaluate a single convenient number, the density weighted true redshift of
the $i$-th photo-z bin. It is defined as 
\be
\langle z_i\rangle=\frac{\int_0^{\infty} zn_i(z)dz}{N_i}=\sum_j
p_{j\rightarrow i}\frac{\int_j zn_i(z)dz}{\int_j n_i(z)dz}\ .
\ee
It is not trivial to robustly evaluate $\langle z_i\rangle$. The special type
of photo-z error, in which the photo-z is a monotonic increasing function of
the true-z, discussed in \S \ref{sec:calibration}, is one of the
problems. However, even if such error does not exist, there is still another
difficulty. We can measure $p_{j\rightarrow i}$ robustly. However, it only
tells us $\int_j  n_i(z)=p_{j\rightarrow i}N_i$. Without proper
parameterization, we can not  derive $n_i$ and thus can not robustly evaluate
$\langle z_i\rangle$.  In reality, for those stage IV lensing surveys, the
function $n_i(z)$ should be sufficiently smooth. This implies that we can
interpolate $p_{j\rightarrow i}$ to model $n_i(z)$. The simulation data that
we used does not have sufficient amount of galaxies allowing us to do this
exercise. Instead, we will try two approximations to evaluate $\langle
z_i\rangle$. 

One possibility is the following approximation
\be
\langle z_i\rangle=\sum_jp_{j\rightarrow i} \langle z^P_j\rangle=\langle z^P_i\rangle+\sum_j(\langle z^P_j\rangle-\langle
z^P_i\rangle)p_{j\rightarrow i}\ ,
\ee
where $\langle z^P_i\rangle$ is the density weighted average photo-z of the
$i$-th  photo-z bin.  The other possibility is to replace $\langle
z^P_i\rangle$ by the middle-point of this redshift bin. Through our simulation
data, we can check the accuracy of the above two approximations. We find that
the first approximation is often better.  The statistical error in $\langle
z_i\rangle$, due to statistical errors in $p_{i\rightarrow j}$ is
then  
\ba
\sigma^2_i&\simeq&\sum_{jk}\left(\langle z^P_j\rangle-\langle z^P_i\rangle)(\langle
z^P_k\rangle -\langle z^P_i\rangle \right)\langle \delta p_{j\rightarrow
  i}\delta p_{j\rightarrow i}\rangle= \sum_{jk}(\langle z^P_j\rangle-\langle
z^P_i\rangle )(\langle z^P_k\rangle -\langle z^P_i\rangle) {\bf
  F}^{-1}_{p_{j\rightarrow i}p_{k\rightarrow i}}\ .
\ea
By far we have assumed that $p_{i\rightarrow j}$ can be any real value. But of
course, $0\leq p_{i\rightarrow j}\leq 1$. This condition can be
straightforwardly enforced by $p\rightarrow (\tanh(x)+1)/2$, where $x$ can be
any real value. Furthermore, there are conditions that
$C^{gg,R}_{ii}>0$, which can be automatically satisfied by the transformation
$C^{gg,R}\rightarrow \exp(y)$, where $y$ can be any real value. Given the
small errors of stage IV lensing projects, these extra
conditions are not likely able to improve the 
reconstruction accuracy significantly. However, for stage 3 projects like
CFHTLS,  these conditions should be explicitly enforced to improve the
reconstruction accuracy. 

\section[]{The induced bias}
\label{sec:appendixB}
The self-calibration is based on the validity of Eq. \ref{eqn:gg} and
\ref{eqn:Gg}. Systematic deviations from these equation then induce
systematic errors in $p_{i\rightarrow j}$. 
Here we outline the calculation of such bias.  The bias in reconstructed
$\lambda$ is 
$\delta \lambda\equiv \lambda^F-\lambda^T$. $\lambda^F$ is determined by
requiring $\partial \ln L/\partial \lambda|_{\lambda^F}=0$.  Taylor expanding
this equation at $\lambda^T$ and keeping terms up to $O(\delta \lambda)$, we
obtain a set of linear equations on $ \Delta \lambda$ and its solution
\be
{\bf F}_{ij} \delta \lambda_j={\bf J}_i\ \ ;\ \ \delta \lambda_i={\bf
  F}^{-1}_{ij}{\bf J}_j\ ,
\ee
where
\ba
\label{eqn:bias}
{\bf J}_m=\sum_{i\leq j} (\sigma^{gg}_{ij})^{-2}\delta
C^{gg}_{ij}\frac{\partial C_{ij}^{gg,P}}{\partial
\lambda_m}+\sum (\sigma^{Gg}_{ij})^{-2}\delta
C^{Gg}_{ij}\frac{\partial C_{ij}^{Gg,P}}{\partial \lambda_m}\ .
\ea
Here, $\delta C^{\alpha}_{ij}\equiv
C_{ij}^{\alpha,D}-C_{ij}^{\alpha,P}(\lambda^T)$ is the model bias in the
corresponding power spectrum.  For example, for the galaxy distribution bias,
we have $ 
\delta C^{gg}_{ij}=\sum_{k}p_{k\rightarrow i}p_{k\rightarrow j}
(\tilde{b}_{k\rightarrow i} \tilde{b}_{k\rightarrow j}-1)C^{gg}_{kk}$ and 
$\delta C^{Gg}_{ij}=\sum_{k>m} p_{k\rightarrow i}p_{m\rightarrow j}
(\tilde{b}_{m\rightarrow j}-1)C^{Gg}_{km}$.  Noticing that, since $\delta
\lambda\propto \delta C$, we have also neglected terms $O(\delta C\times \delta
\lambda)$ when deriving this equation. All differentials in Eq. \ref{eqn:bias}
are evaluated at $\lambda^T$. This result agrees with that in
\citet{Huterer05,Huterer06}.  The related matrices are 
calculated from the following expressions,
\ba
{\bf J}_{p_{\mu\rightarrow \nu}}=\sum_{j\geq \nu} (\sigma^{gg}_{\nu j})^{-2}
\delta C^{gg}_{\nu j} A^{gg}_{\mu\nu j}+\sum_{i\leq \nu} (\sigma^{gg}_{i\nu})^{-2}
\delta C^{gg}_{i\nu} B^{gg}_{\mu\nu i}+\sum (\sigma^{Gg}_{\nu j})^{-2}
\delta C^{Gg}_{\nu j} A^{Gg}_{\mu\nu j}+\sum (\sigma^{Gg}_{i\nu})^{-2}
\delta C^{Gg}_{i\nu} B^{Gg}_{\mu\nu i}\ ,
\ea
\ba
{\bf J}_{C^{Gg}_{\mu \nu}}=\sum (\sigma^{Gg}_{ij})^{-2}\delta
C^{Gg}_{ij}p_{\mu\rightarrow i}p_{\nu\rightarrow j}\ ,
\ea
and 
\ba
{\bf J}_{C^{gg}_{\mu \nu}}=\sum_{i\leq j} (\sigma^{gg}_{ij})^{-2}\delta
C^{gg}_{ij}p_{\mu\rightarrow i}p_{\nu\rightarrow j}\ .
\ea
The induced bias in individual $p_{i\rightarrow j}$ is shown in
Fig. \ref{fig:bias}. 
The induced bias in the mean redshift is
\be
\delta z_i\simeq \sum_j(\langle z^P_j\rangle-\langle z^P_i\rangle)\delta
p_{j\rightarrow i}\ .
\ee
The numerical results are shown in Fig. \ref{fig:zmeanbias} \&
\ref{fig:zmeanbias22}.

\end{document}